\documentclass[twocolumns]{aa}
\usepackage{graphicx}
\usepackage{amssymb}
\usepackage{rotate}
\begin{document}

\title{On the fine structure of sunspot penumbrae\\
II. On the nature of the Evershed flow}

\author{Borrero, J.M.\inst{1}\thanks{\emph{Present address:} 
High Altitude Observatory, 3450 Mitchell Lane, 80301 Boulder, 
Colorado, USA. Email: borrero@ucar.edu} \and Lagg, A.\inst{1}
\and Solanki, S.K\inst{1} \and Collados, M.\inst{2}}
\institute{Max--Planck Institut f\"ur Sonnensystemforschung, 37191,
Katlenburg--Lindau, Germany \and Instituto de Astrof{\'\i}sica
de Canarias, E-38200, V{\'\i}a L\'actea s/n, La Laguna, 
Tenerife, Spain}

\abstract{We investigate the fine structure of
the sunspot penumbra by means of a model that
allows for a flux tube in horizontal pressure balance with the
magnetic background atmosphere in which it is embedded.
We apply this model to spectropolarimetric observations
of two neutral iron lines at 1.56 $\mu$m and invert several radial
cuts in the penumbra of the same sunspot at two different heliocentric
angles. In the inner part of the penumbra we find hot flux tubes 
that are somewhat inclined to the horizontal. They become gradually 
more horizontal and cooler with increasing radial distance. This is 
accompanied by an increase in the velocity of the plasma and a
decrease of the gas pressure difference between flux tube and
the background component. At large radial distances the flow speed
exceeds the critical speed and evidence is found for the formation
of a shock front. These results are in good agreement with
simulations of the penumbral fine structure and provide strong 
support for the siphon flow as the physical mechanism driving 
the Evershed flow.
}

\authorrunning{Borrero et al}
\titlerunning{On the nature of the Evershed flow}
\maketitle

\section{Introduction}%

The picture of the magnetic fine structure of the penumbra has strongly
evolved over the last decade (e.g., Degenhardt \& Wiehr 1991; Title et
al. 1993; Solanki \& Montavon 1993; Rimmele 1995; Stanchfield et al. 1997; 
Westendorp Plaza et al. 1997;
Schlichenmaier et al. 1998a; Scharmer et al. 2002; see Solanki 2003, Bellot Rubio 2003 
and Thomas \& Weiss 2004 for an overview). It is now
accepted that the penumbral magnetic field is {\it uncombed}, i.e. 
inclined at least in two different directions on a small scale. 
There is also considerable evidence that the more horizontal component
must be in the form of flux tubes, although the diameter of these flux
tubes is still a matter of debate (S\'anchez Almeida 1998, 2001;
Mart{\'\i}nez Pillet 2000, 2001). These flux tubes carry the Evershed flow
(Evershed 1909; Title et al. 1993; Westendorp Plaza et
al. 2001a,2001b; Bellot Rubio et al. 2003,2004; Borrero et al. 2004). 
Many of the tubes return to the solar interior within 
the penumbra (Westendorp Plaza et al. 1997; del Toro Iniesta et
al. 2001; Mathew et al. 2003; Borrero et al. 2004) and along
with the magnetic flux a large fraction of the mass flux carried
by the Evershed flow returns also to the solar interior within
the penumbra or just outside it (B\"orner \& Kneer 1992; Solanki et
al. 1994, 1999). 

The combination of these results raises questions
regarding the commonly considered physical mechanism driving the
Evershed flow. Since the wave hypothesis is ruled out (B\"unte \&
Solanki 1995) and episodic Evershed flow produced when a flux tube
{\it falls} and drains (Wentzel 1992) faces difficulties due to
the relative immutability of the penumbral magnetic pattern 
(Solanki \& R\"uedi 2003), the most widely accepted mechanism is that
the flow is caused by a gas pressure gradient between the upflowing
and the downflowing footpoints, also referred to as the siphon flow mechanism 
(Meyer \& Schmidt 1968; see also Thomas 1988). Commonly, this
pressure gradient is thought to be produced by a difference in the
field strength between the footpoints which, due to horizontal
pressure balance, leads to a gas pressure difference (e.g. Degenhardt
1989,1991; Thomas \& Montesinos 1991,1993, Montesinos \& Thomas 1997). All else being equal, the gas
flows from the footpoint with lower field strength to the footpoint
with the higher field strength. However, if most of the gas flows
only within the penumbra, then due to the roughly factor of 2 larger 
magnetic field strength at the inner boundary of the penumbra compared
to the outer edge, one would naively expect the gas to flow inwards,
contradicting observations. Montesinos \& Thomas (1997) have argued
that this radial decrease of the field strength is only apparent,
being caused by different $\tau=1$ levels at the footpoints and the fact that local 
intense magnetic flux concentrations at the outer penumbral edge could not be easily resolved.

A possible resolution of this dilemma was noticed by Borrero et al
(2004; hereafter Paper I), who
found that whereas the strength of the inclined magnetic component
drops very rapidly in the radial direction, as required by a global magnetic structure close to
potential (see Jahn 1989), the 
horizontal component carrying the Evershed flow shows far less 
variation (cf. R\"uedi et al. 1998, 1999).
The analysis in Paper I was incomplete in the sense that the two
components were independent of each other. Here we overcome this
shortcoming and take into account the flux-tube structure of the
field and the pressure balance between components.
 We apply a powerful inversion technique (described in Section 2) to
spectropolarimetric observations of infrared Zeeman sensitive
spectral lines (Section 3). General results are presented
and discussed in Section 4. In Section 5 we discuss the generation
of Net Circular Polarization by the uncombed model and the
implications for the typical size of the penumbral flux tubes.
Section 6 describes our results in the framework
of the theoretical models employed to explain the Evershed effect
in terms of gas pressure differences. Our main findings and
conclusions are summarized in Section 7.

\section{Description of the penumbral model and Stokes profile inversion}%

The analysis of spectropolarimetric observations of the sunspot
penumbra by means of Stokes profile inversion has, so far, either 
considered 2 distinct components (assuming the physical 
quantities to be constant with depth: Bellot Rubio et al. 2003,2004) 
or 1 component (allowing gradients to be present; Westendorp Plaza et
al. 1997, 2001a, 2001b; Bellot Rubio et al. 2002; Mathew et al. 2003).
In Paper I we suggested that the {\it uncombed}
penumbral model described by Solanki \& Montavon (1993) provides
a picture for the penumbral fine structure that is able to encompass the results
of these investigations. This conclusion was based on the application 
of a Stokes profile inversion technique assuming the two different
geometries mentioned above, together with
considerations on how the area asymmetry of the circular polarization
profiles is influenced by gradients in the magnetic and kinematic 
stratifications.

In this paper we carry out Stokes profile inversions based on the uncombed
model. This consists of a flux tube embedded in a magnetic
surrounding atmosphere.  The forward modelling for the considered geometry has 
already been addressed (see e.g. Degenhardt \& Kneer 1992, Solanki
\& Montavon 1993; Mart{\'\i}nez Pillet 2000).
The basic geometry is illustrated in
Figure 1. The simplest representation of the uncombed model
is in terms of two rays: the first ray passes
along the surrounding atmosphere (vertical dot-dashed line pointing
towards the observer) while the other ray cuts both surrounding atmosphere
and the flux tube (vertical dashed
line)\footnote{Note that this geometrical simplification in terms of
  two rays implies that we are considering the cross section of the flux
  tube to be square instead of circular as in Fig.~1.}.
Let us denote with {\bf $\chi_{\rm s}$} any of the magnetic and kinematic 
physical quantities for the surrounding atmosphere\footnote{{\bf $\chi_{\rm s}$} is
  assumed to be height independent.}. For the flux tube 
component we adopt the following form for {\bf $\chi_{\rm t}(z)$}:

\begin{equation}
\chi_{\rm t}(z) = \left\{ \begin{array}{ll}
  \chi_{\rm t} & \rm{if} \;\;\; z \in [z_{0}-R_{\rm t},z_{0}+R_{\rm t}]\\
  \chi_{\rm s} & \rm{otherwise}
  \end{array} \right.
\end{equation}

\noindent where $\chi_{\rm t}$ on the right hand side of Eq.~1 is height
independent, $z_0$ is the height where the axis of the flux tube is
located and $R_{\rm t}$ is its radius.
Note that Eq.~1 implies that physical stratifications along
the flux tube atmosphere (vertical dashed line in Fig.~1) are the
same as in the surrounding atmosphere above and beneath the
flux tube: $z<z_o-R_{\rm t}$ and $z>z_o+R_{\rm t}$. At the flux tube lower
and upper boundaries ($z^{*}=z_0 \pm R_{\rm t}$) the physical quantities suffer a jump
whose magnitude is $\Delta \chi=\chi_{\rm t}-\chi_{\rm s}$.
These jumps/gradients are the essential ingredients to explain the 
Net Circular Polarization (NCP)
observed in the sunspot penumbra (Solanki \& Montavon 1993; cf.
Mart{\'\i}nez Pillet 2000).

\begin{figure}
\begin{center}
\includegraphics[width=7cm]{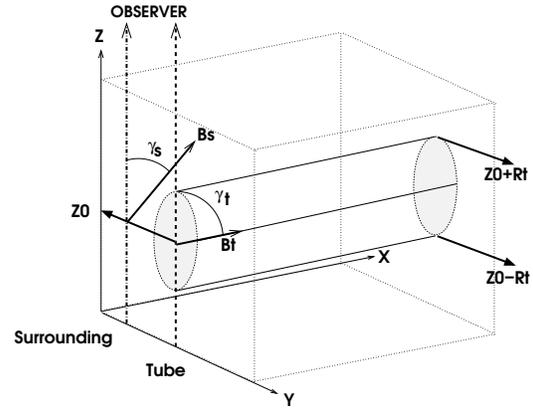}
\end{center}
\caption{Example of the geometrical scheme used in the inversion. 
The radiative transfer equation is solved along the
2 rays (dashed and dot-dashed lines) representing
the embedded flux tube and surrounding atmosphere respectively.
$\gamma_{\rm s}$ and $\gamma_{\rm t}$ refer to the inclination of
the magnetic field vector with respect to the observer.
In this picture for simplicity the heliocentric angle
is $\theta=0$ and $\gamma_{\rm t}=90^{\circ}$.}
\end{figure}


The inversions based on this geometry have been carried out using the
inversion code SPINOR (see Frutiger et al. 1999; Frutiger 2000).
The code performs spectral line synthesis in Local Thermodynamic
Equilibrium (LTE) and employs a Levenberg-Marquart nonlinear
least-squares $\chi^2$ minimization algorithm (Press et al. 1986), whereby derivatives are
calculated through numerical response functions (RFs; see Ruiz Cobo
\& del Toro Iniesta 1992). The free parameters allowed in the
inversion are, for the surrounding atmosphere: $V_{\rm LOS,s}$ 
(line of sight velocity), $B_{\rm s}$ (magnetic field strength), 
$\gamma_{\rm s}$ (magnetic field inclination with respect to the 
observer), $\phi_{\rm s}$ (magnetic field azimuth) and $T_{\rm
  s}(\tau_5=1)$ for temperature, where $\tau_5$ is the optical depth at a
reference wavelength of 5000 \AA. The height dependence of $T_{\rm
  s}(\tau)$ is taken from the penumbral model by del Toro 
Iniesta et al. (1994). When $T_{\rm s}(\tau_5=1)$ is changed by $\Delta T$
the temperature at all heights is changed additively: $T_{\rm s}(\tau)=T_{\rm
  s,old}(\tau)+\Delta T$.

For the flux tube the free parameters are: 
$V_{\rm LOS,t}$, $B_{\rm t}$, 
$\gamma_{\rm t}$, $\phi_{\rm t}$, macro and microturbulent velocities,
$v_{\rm mac,t}$ and $v_{\rm mic,t}$, and $T_{\rm t}(\tau_5=1)$ 
(where again $T_{\rm t}(\tau)$ is taken from the mean penumbra model of 
del Toro Iniesta et al.). In addition, $z_0$, $R_{\rm t}$ and 
$\alpha_{\rm t}$ (fractional horizontal area covered by
the flux tube component with respect to the total area) are also
allowed to change. Finally we employ a stray light correction to model
the contribution of light from the neighbouring granulation that
enters into the spectrograph's slit. To this end we used the two
component model for the quiet sun from Borrero \& Bellot Rubio (2002)
to produce synthetic intensity profiles, $I_{\rm q}$, of the observed 
spectral lines (see Sect.~3) and combined it with the emergent
spectrum of the pure penumbral profiles using a filling factor
$\alpha_{\rm q}$ which is also retrieved from the inversion.
This results in a total of 16 free parameters. This compares positively with
the inversions carried out in Paper I, which used a total number of 23 and 18 free
  parameters for the 1 component and 2 component inversions, respectively.

The radiative transfer equation is integrated using the Hermitian
Approach (see Bellot Rubio et al. 1998) for each ray separately.
The Stokes profiles from quiet sun, flux tube and surrounding
atmosphere are finally combined using the filling factors 
$\alpha_{\rm t}$ and $\alpha_{\rm q}$:

\begin{eqnarray}
{\textbf{\textit{S}}}(\lambda) &=& \alpha_{\rm q}
{\textbf{\textit{S}}}_{\rm q}(\lambda) + 
(1-\alpha_{\rm q})[(1-\alpha_{\rm t}) {\textbf{\textit{S}}}_{\rm s}(\lambda) + 
\alpha_{\rm t} {\textbf{\textit{S}}}_{\rm t}(\lambda)] \;\;\; ,
\end{eqnarray}

\noindent where ${\textbf{\textit{S}}}$ represents the Stokes vector (I,Q,U,V): 
as arising from the surrounding
atmosphere (dot-dashed ray in Fig.~1) ${\textbf{\textit{S}}}_{\rm s}$,
 from the ray cutting the flux tube (dashed ray in Fig.~1)
${\textbf{\textit{S}}}_{\rm t}$, as well as the quiet sun contribution
${\textbf{\textit{S}}}_{\rm q}=(I_{\rm q},0,0,0)$ which is a non polarized
contribution and only affects the total intensity profiles.
These synthetic profiles are now compared with the observations 
and the free parameters are changed according to the RF until 
the minimum $\chi^2$ is reached. A detailed study of the numerical performance 
of this procedure as well as the uniqueness of the retrieved
atmosphere under different levels of noise in the observations is 
presented by Borrero et al. (2003a).

The radiative transfer is always performed in the optical depth scale,
but the flux tube and its force balance with its surroundings are more
naturally described in the geometrical depth scale. The correct
functioning of this interplay requires a sufficiently complex
procedure. First, a geometrical height
scale is assigned to the surrounding atmosphere following the strategy outlined 
in Gray (1992) and integrating the hydrostatic equilibrium equation (assuming
a force free situation $\nabla \times {\bf B} \parallel 
{\bf B}$),

\begin{equation}
\frac{\partial P_{g,s}}{\partial \tau_s} = \frac{g}{\kappa_{c,s}}
\end{equation}

\noindent where $g$ is the solar surface gravitational acceleration and $\tau$ is
the optical depth computed using the continuum opacity, $\kappa_c$,
at 5000 \AA. This requires an estimate of the gas
pressure at the top of the tabulated atmosphere: $P_{g,s}(\tau_{s,max})$.
With this, the gas pressure stratification $P_{g,s}(\tau_s)$ 
is obtained in the surrounding atmosphere, and since the temperature is
obtained from the inversion, the equation of state (ideal gas law
including a variable mean molecular weight to account for the ionization
of the different species) provides the density: $\rho_s(\tau_s)$.
The relation $d\tau_s= -\rho_s \kappa_{c,s} dz$ is now integrated
setting $z=0$ at $\log\tau_5=0$ and thus defining the geometrical
height scale.

The gas pressure in the tube component is obtained under the
assumption of total pressure balance with the surroundings
at the height of the axis of the flux tube, 

\begin{equation}
P_{\rm g,t}(z) = \left\{ \begin{array}{ll}
P_{\rm g,s}(z) + \frac{(B_{\rm s}^{2}-B_{\rm t}^{2})}{8\pi} & {\rm if} \;\; z \in
[z_{0}-R_{\rm t},z_{0}+R_{\rm t}]\\
P_{\rm g,s}(z) & \rm{otherwise}
\end{array} \right.
\end{equation}

This assumption is valid as long as the magnetic field of the external 
atmosphere does not penetrate into the flux tube and vice versa 
(Kippenhahn \& M\"ollenhof 1975, Chap.3).
Since $P_{g,t}(z)$ is now known and $T_{t}(z)$ was obtained
from the inversion, the density $\rho_t(z)$ can be evaluated by using the idal
gas equation, and thus
a new optical depth scale for the atmosphere containing the flux tube
can be obtained through the relation: $d\tau_t=-\rho_t \kappa_{c,t}
dz$. For the integration of this last equation a boundary condition
is employed which implies that for $z > z_0+R_{\rm t}$ the surrounding
and flux tube components must have the same $z$ values: 
$z(\tau_{t})=z(\tau_{s})$. Note that the obtained flux tube density, $\rho_t(z)$,
does not satisfy vertical hydrostatic equilibrium.

\section{Observations and data reduction}%

The active region NOAA AR 8706 was observed
on Sep 21st, 1999 and Sep 27th, 1999 at $\mu=cos\theta=0.51$ 
and $\mu=0.91$ respectively, using TIP (Tenerife Infrared Polarimeter, 
Mart{\'\i}nez Pillet et al. 1999) attached to the spectrograph 
of the 70cm German VTT of the Observatorio del Teide. Here, $\theta$ is the
heliocentric angle. The recorded 
spectral region contains the full polarization profiles of the pair 
of Fe I lines 15648.5 \AA~ ($g=3$) and Fe I 15652.8 \AA ($g_{\rm eff}=1.53$). 
The wavelength sampling is about 29 m\AA. 
The diagnostic properties of these lines
have been discussed by Solanki et al. (1992). They are formed in the deep
photosphere as a result of their high excitation potentials and the
low continuum opacity at this wavelength. These lines sample
a relatively narrow layer not wider than $\log\tau_5=[0.5,-2]$
(see Bellot Rubio et al. 2000; Mathew et al. 2003). In the umbra
the second neutral iron line is heavily blended by two molecular OH lines at
15651.9 and 15653.5 \AA. These lines are Zeeman sensitive and belong to the Meinel
system 3,1 (see Berdyugina \& Solanki 2002 for details). Their equivalent
widths greatly decrease towards the penumbra. However they can still
be seen clearly at the umbra-penumbra boundary. In order to accurately analyze
both atomic lines we also compute the OH lines. The amployed atomic parameters 
for the observed Fe I lines were taken from Borrero et al. (2003b), 
while for the OH lines the values given by Abrams et al. (1994) and Mies (1974) 
were used (see Table 1).

The usual data reduction procedures for TIP were in general followed; 
we proceeded carefully at several points, however. Firstly the neutral iron
line, Fe I 15648.5, appears to be blended by a telluric line in its
red wing. This complication was overcome by fitting the average quiet
Sun profile of the second line with an appropriate model atmosphere.
This model was then used to synthetize the first iron line.
The ratio between the average observed and computed line allowed us
to recover the shape of the telluric blend, 
which was subsequently eliminated from the rest of the profiles. 
This procedure, although not perfect, does not introduce modifications 
to the original profile (equivalent width and line core intensity) larger than
$\sim$3 \%. 

Secondly, the continuum correction posed a considerable problem.
The four quadrant configuration of the TIP camera produce small
gradients in the continuum intensity that remain after applying
the flat field correction. To account for this we compared our
spatially averaged flat field with the infrared FTS atlas from Livingston \&
Wallace (1991) and defined several wavelength positions where the continuum
should be reached. A smooth second order polynomial is interpolated
over those selected points and used to bring the continuum in the
flat field image to the same level as the FTS continuum.
Thirdly, the polarization signals were corrected for residual cross talk 
using the statistical approach described in Collados (2001; see also
Schlichenmaier \& Collados 2002). Finally, the wavelength was calibrated 
by assigning the laboratory wavelengths to the cores 
of the Fe I lines in the average quiet Sun profiles. To account for
the effects of the granulation we shifted this wavelength scale
by $-$400 m s$^{-1}$, which is the approximate
value for the convective blueshift as deduced from
the quiet sun model of Borrero \& Bellot Rubio (2002). Second order
  corrections due to the different viewing angles (Balthasar 1985) have not
been performed.

The seeing conditions were rather good
during the observations, with the granulation being clearly
discernible in the reconstructed continuum images 
(see Figure 2; left panels). By calculating the power spectrum 
of the continuum intensity in the neighbouring quiet Sun we estimate 
the spatial resolution to be about 1 arc sec. In Fig.~2 (right panels) 
the maps of the total circular polarization are also shown.
Due to projection effects the polarity inversion line
(region in the limb side of the penumbra where the Stokes $V$ signal
changes its sign, i.e., where the average magnetic field is perpendicular to
the observer) appears in the outer penumbra when the sunspot is observed 
near disk center ($\mu=0.91$) but lies closer to the umbra
when the spot is located closer to the limb ($\mu=0.51$). Note that the arrows
in Fig.~2 point towards the solar disk center. Part of this data set 
has been analyzed in Paper I (see also Mathew et al. 2003, 2004).

\begin{figure*}
\begin{center}
\begin{tabular}{cc}
\includegraphics[width=6cm]{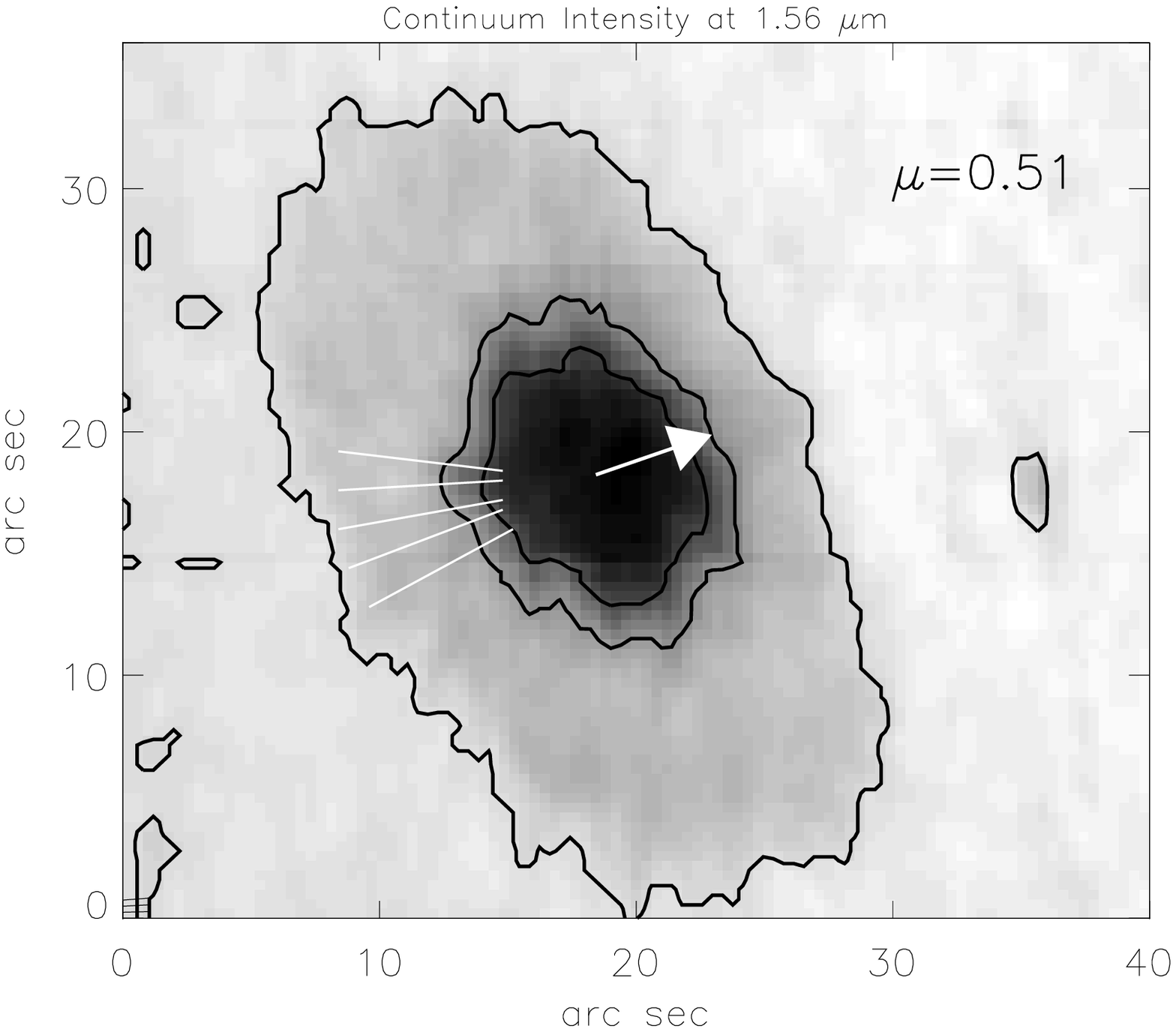} &
\includegraphics[width=6cm]{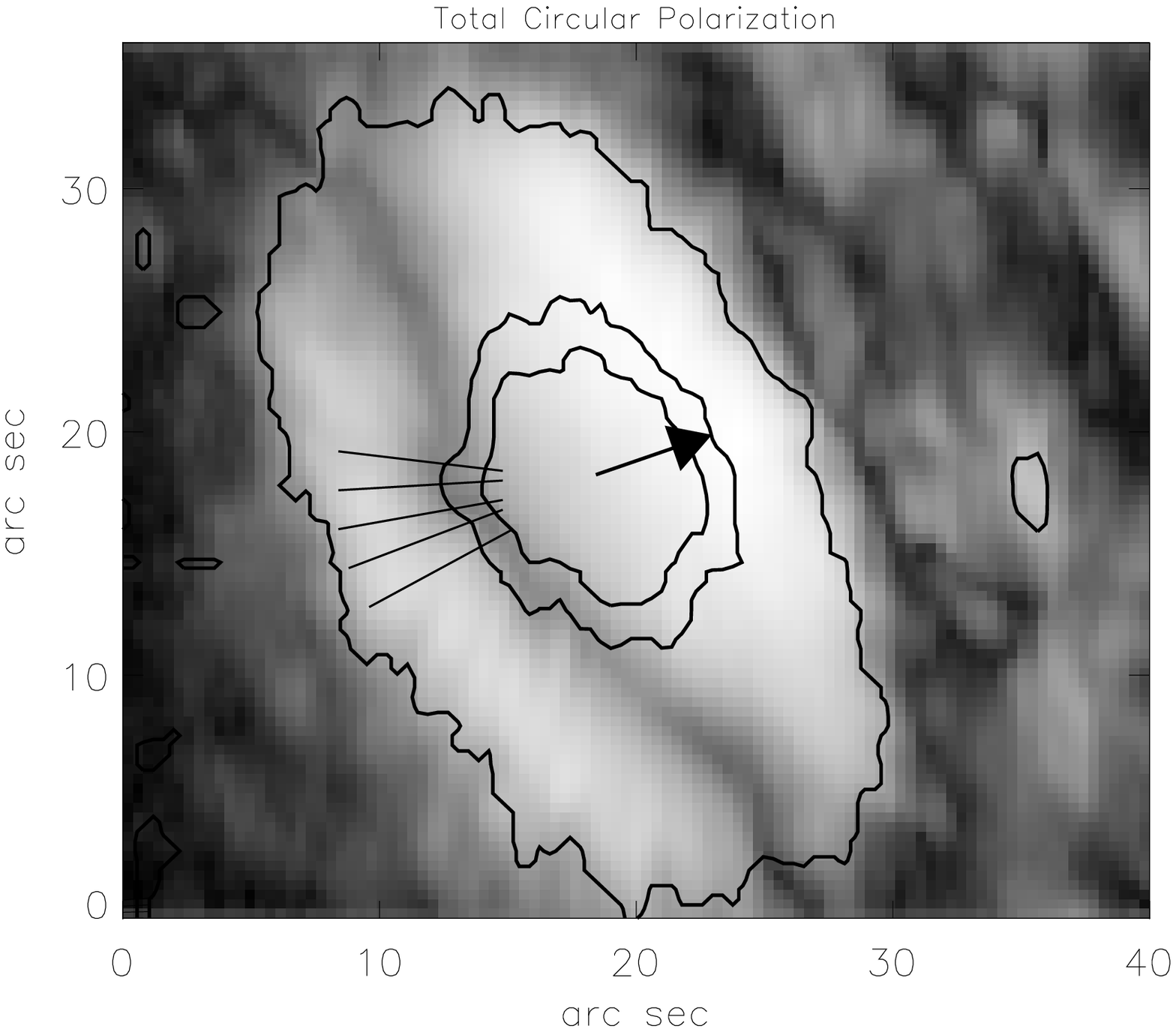} \\
\includegraphics[width=6cm]{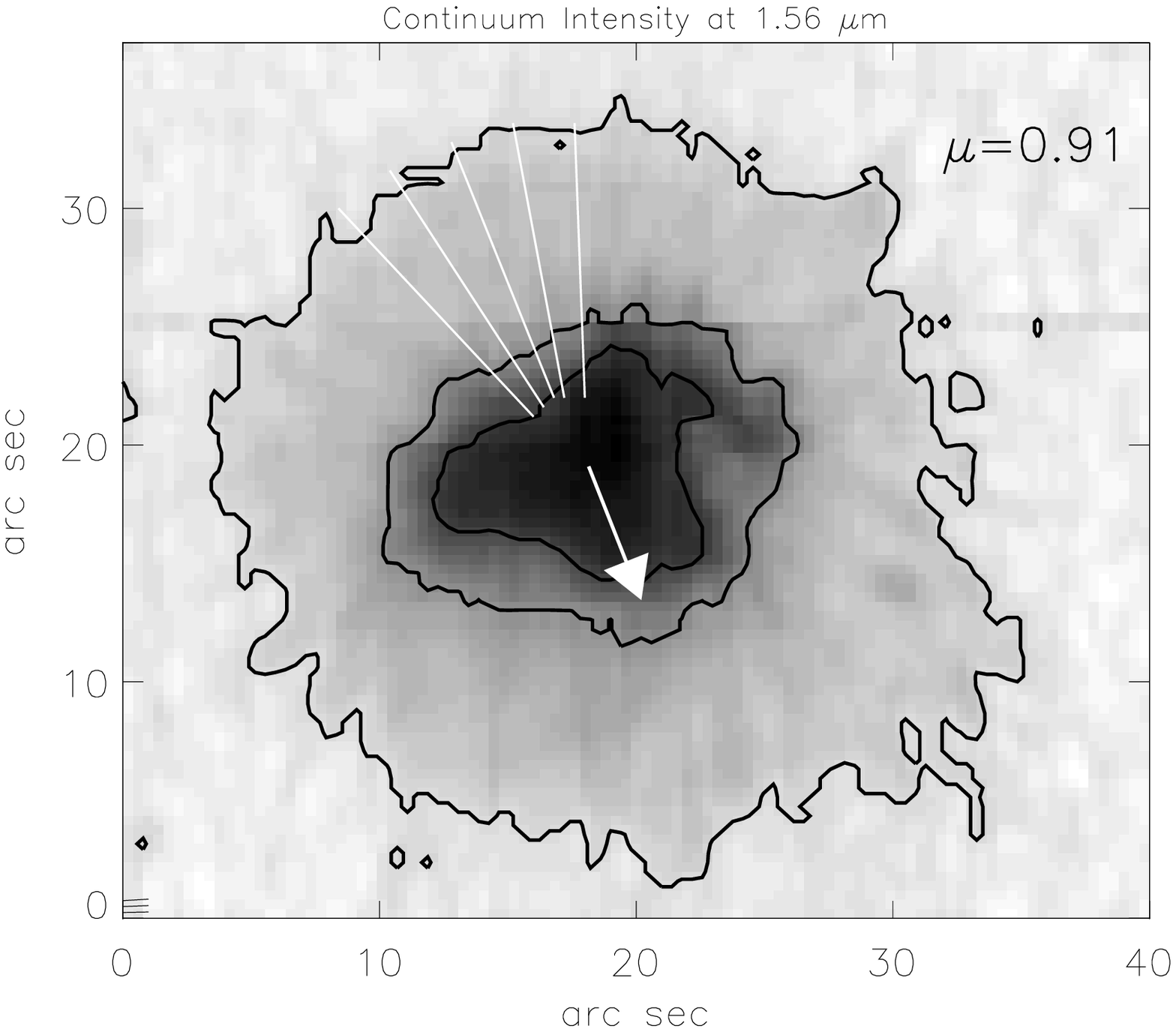} &
\includegraphics[width=6cm]{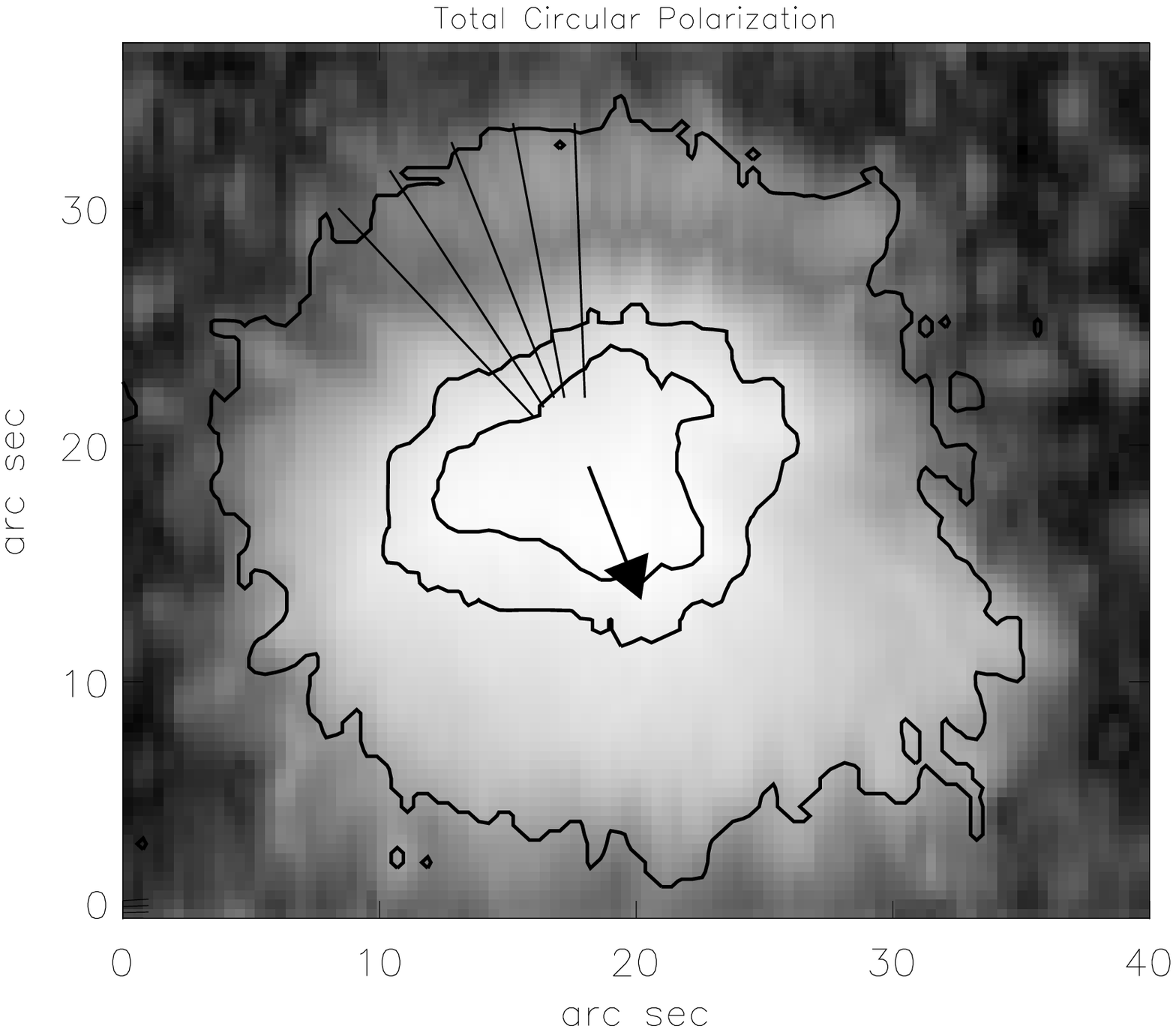}
\end{tabular}
\end{center}
\caption{Top panels: NOAA 8706 observed in 21 September 1999 at
a heliocentric angle $\mu=0.51$ (left: continuum intensity map
at 1.56 $\mu$m; right: total circular polarization map for Fe I
15648.5 \AA). Bottom panels: NOAA 8706 observed in 27 September 1999 at a
heliocentric angle $\mu$=0.91 (left: continuum intensity map
at 1.56 $\mu$m; right: total circular polarization map for Fe I
15648.5 \AA). The arrows point towards the direction of the solar
disk center. The two inner-most contours enclose the
umbral-penumbral boundary. The external contour defines the sunspot radius
$r=R$, at each position angle. These three contours have been defined
as $0.45 I_{\rm c}$,$0.65 I_{\rm c}$ and $0.85 I_{\rm c}$, where
$I_{\rm c}$ represents the average continuum intensity of the quiet Sun.
The radial cuts selected for our analysis are also shown.}
\end{figure*}

\begin{table}
\caption[]{Atomic and molecular parameters of the observed spectral
  lines. $\lambda_0$ 
represents the laboratory central wavelength, $\chi_{\rm l}$ the
excitation potential of the lower energy level, and $\log gf$ the 
logarithm of the oscillator strength times the multiplicity of
the level. The parameters $\alpha$ and $\sigma$ (in units of 
Bohr's radius, $a_0$) are used to calculate the broadening of
the lines by collisions with neutral hydrogen atoms as resulting 
from the ABO theory (Barklem \& O'Mara 1997). The last column gives 
the effective Land\'e factor of the transition, $g_{\rm eff}$. For the molecular
lines $I_{\rm U}$, $I_{\rm L}$, $V_{\rm U}$ and $V_{\rm L}$ represent
the upper/low multiplets sublevels and vibrational levels
respectively. $J_{\rm L}$ stands for the rotational number of the
lower level. Finally, the oscillator strength of the molecular transition
  is given.}
\begin{center}
\tabcolsep .6em
\begin{tabular}{lccccccccc}
\hline
\hline
Atom & $\lambda_{0}$ & $\chi_{l}$ & log~gf & $\alpha$ &
$\sigma$ & $g_{\rm eff}$ \\
& (\AA~) & (eV) & (dex) & & ($a_{0}^{2}$)\\\hline
\hline
Fe I & 15648.515 &     5.426 &    $-$0.675  &  0.229  &     977 & 3.00\\           
Fe I & 15652.874 &     6.246 &    $-$0.043  &  0.330  &    1444 & 1.53\\
\hline
\hline
Molecule & $\lambda_{0}$ & Branch & $I_{\rm U}$-$I_{\rm L}$ & $V_{\rm
  U}$-$V_{\rm L}$ & $J_{\rm L}$ & $f$\\
\hline
OH & 15651.895 & P & 1-1 & 3-1 & 6.5 & 0.8$\times$10$^{-6}$\\
OH & 15653.478 & P & 1-1 & 3-1 & 6.5 & 0.8$\times$10$^{-6}$\\
\hline
\end{tabular}
\end{center}
\end{table}

\section{Results and discussion}%

We have inverted individually all the pixels along the radial cuts 
shown in Fig.~2 (5 for each heliocentric angle). Each cut contains approximatively 20
pixels and ranges roughly from $r/R=0.4$ to $r/R=0.9$ (where $R$
is the penumbral radius, indicated by the external contour in Fig.~2).
We have chosen the cuts such that
they all lie on the limbward side of the penumbra and near the line
of symmetry (i.e. the line connecting the sunspot's center and the center
of the solar disk), since this is where the flux tubes 
leave the most distinctive fingerprints on the observed profiles, 
thus allowing for a reliable determination of their properties, as shown in
Paper I.

\subsection{Example}
In Fig.~3 we present an example of the observed (filled circles) 
and fitted (solid lines) circular polarization profiles for a penumbral
point. These multi-lobed profiles can be successfully fitted by adding
together two Stokes $V$ profiles. The profile from the
surrounding atmopshere, $V_{\rm s}$, is produced by a positive polarity magnetic 
field ($\gamma < 90^{\circ}$; dot-dashed line) with zero line-of-sight 
velocity (zero crossing is at $\lambda_{0}$), and the Stokes $V$ profile
from the ray cutting through the flux tube, $V_{\rm t}$, which is produced 
by a negative polarity magnetic field ($\gamma > 90^{\circ}$; dashed
line) carrying a flow directed away from the observer
(zero crossing is red shifted with respect to $\lambda_{0}$).
The filling factor of the flux tube atmosphere $\alpha_{\rm t}$
and of the stray light contribution $\alpha_{\rm q}$ are applied
to $V_{\rm s}$ and $V_{\rm t}$ to obtain the final
emergent profile (solid line) according to Eq.~2.

In Fig.~4 we present the atmospheric stratifications resulting 
from the inversion of the profiles shown in Fig.~3 using the
uncombed model (see Sect.~2). Outside the flux tube the magnetic field
strength is about 2000 Gauss and is inclined with respect to the
observer by about 75$^{\circ}$. Inside the flux tube the magnetic field
is weaker ($\simeq 1300$ Gauss) and more inclined ($\gamma \simeq
100^{\circ}$). While the surrounding atmosphere is basically at rest, 
$V_{\rm LOS,s} \simeq 0.2$ km~s$^{-1}$, we
detect red shifts of $V_{\rm LOS,t} \simeq 1.1$ km~s$^{-1}$ inside the flux
tube. In addition, the results from the inversion indicate that the flux tube
is hotter, by roughly 500 K, than the surrounding atmosphere.
The presence of the discontinuity along the flux tube atmosphere
produces the area asymmetry in Stokes $V$, $\delta A_{\rm FIT}$, 
marked in Fig.~3.

\begin{figure*}
\begin{center}
\begin{tabular}{cc}
\includegraphics[width=7.5cm]{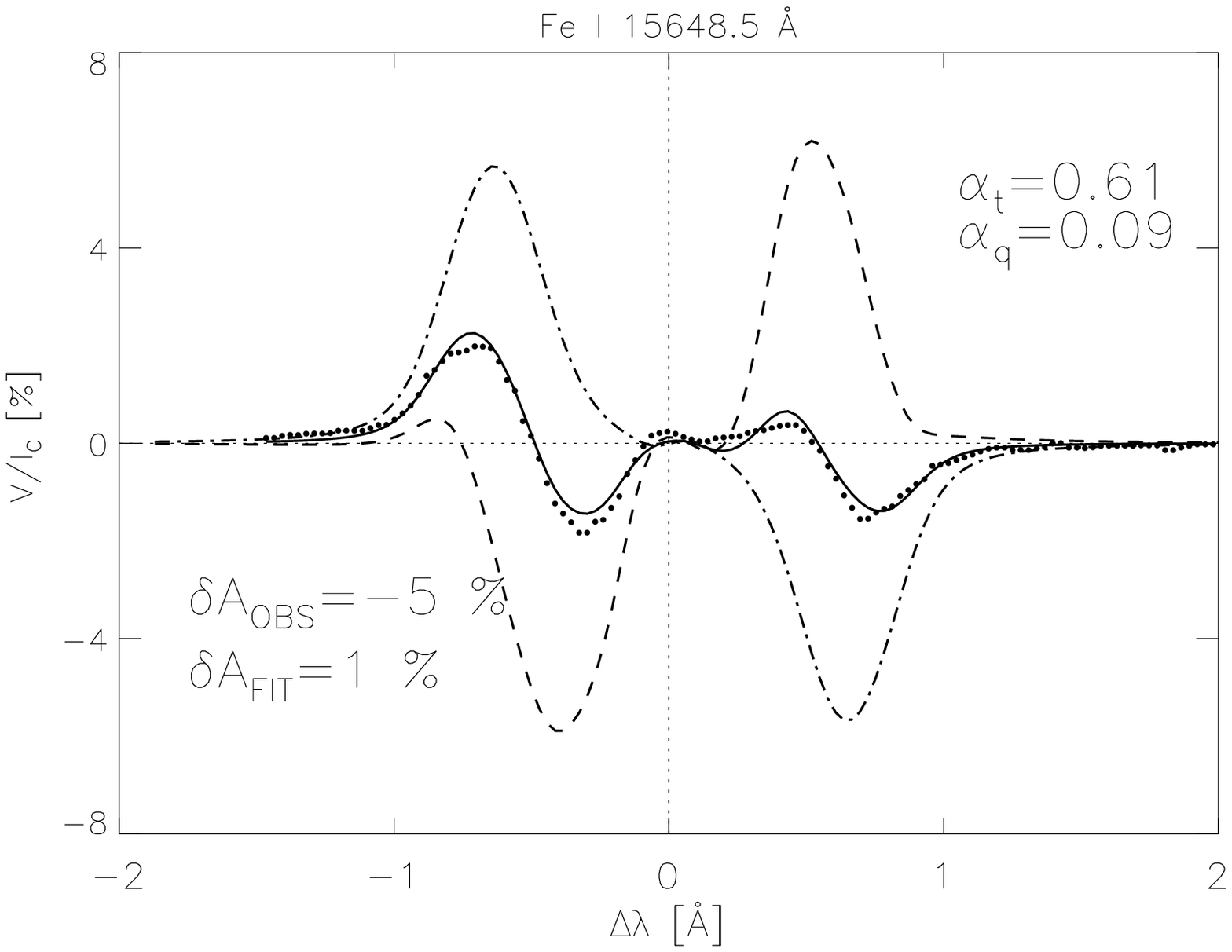} &
\includegraphics[width=7.5cm]{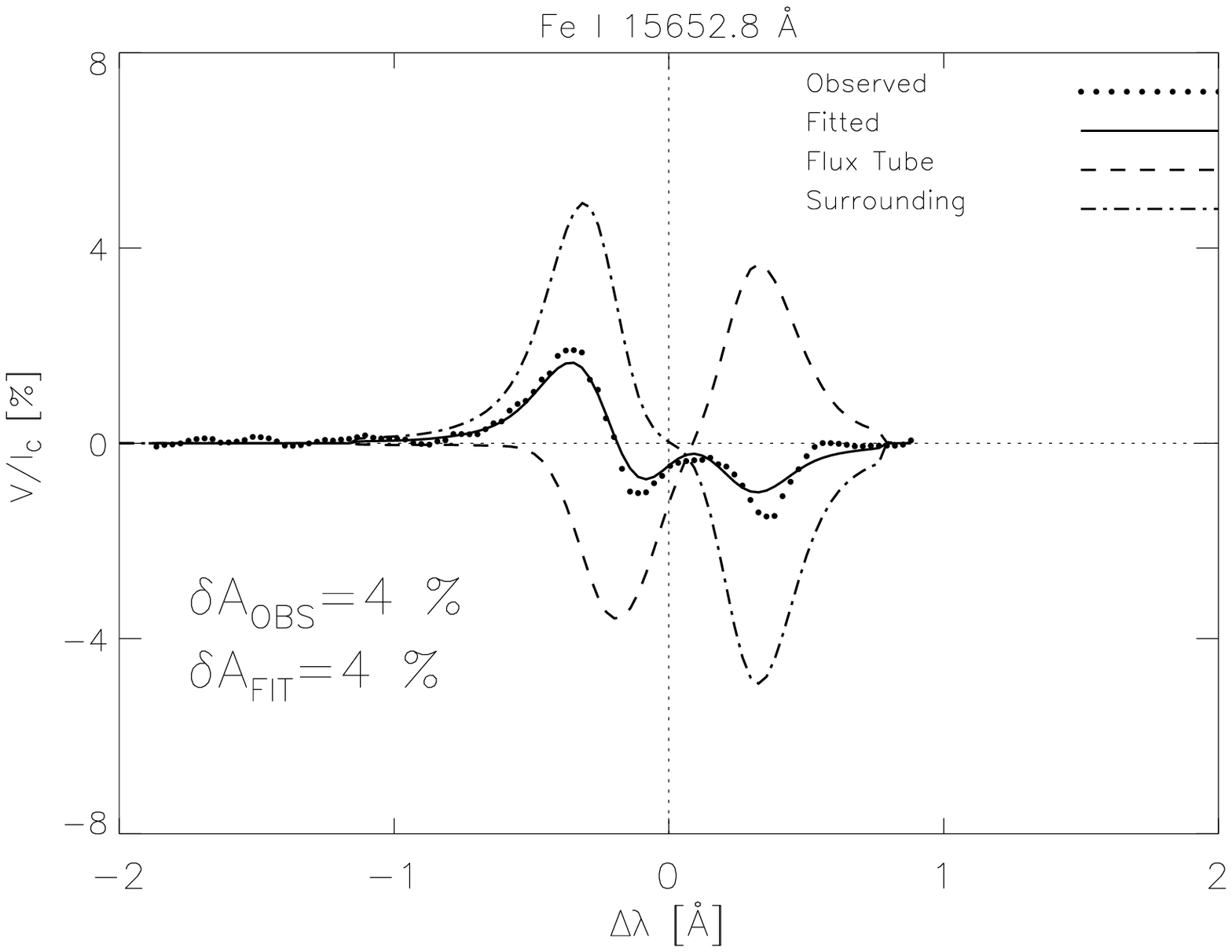} \\
\end{tabular}
\end{center}
\caption{Example of the observed ({\bf dots}) and fitted (solid
lines) Stokes $V$ profiles (left: 15648.5 \AA; right: 15652.5 \AA)
for a penumbral point. The fitted profile is obtained by
linearly combining the profile emerging from the surrounding component
(dot-dashed line) and the profile
emerging from the ray piercing the flux tube (dashed line). 
The employed filling factors,
$\alpha_{\rm t}$ and $\alpha_{\rm q}$, as well as the area
asymmetry of the observed and fitted profiles are also given.}
\end{figure*}

\begin{figure*}
\begin{center}
\includegraphics[width=15cm]{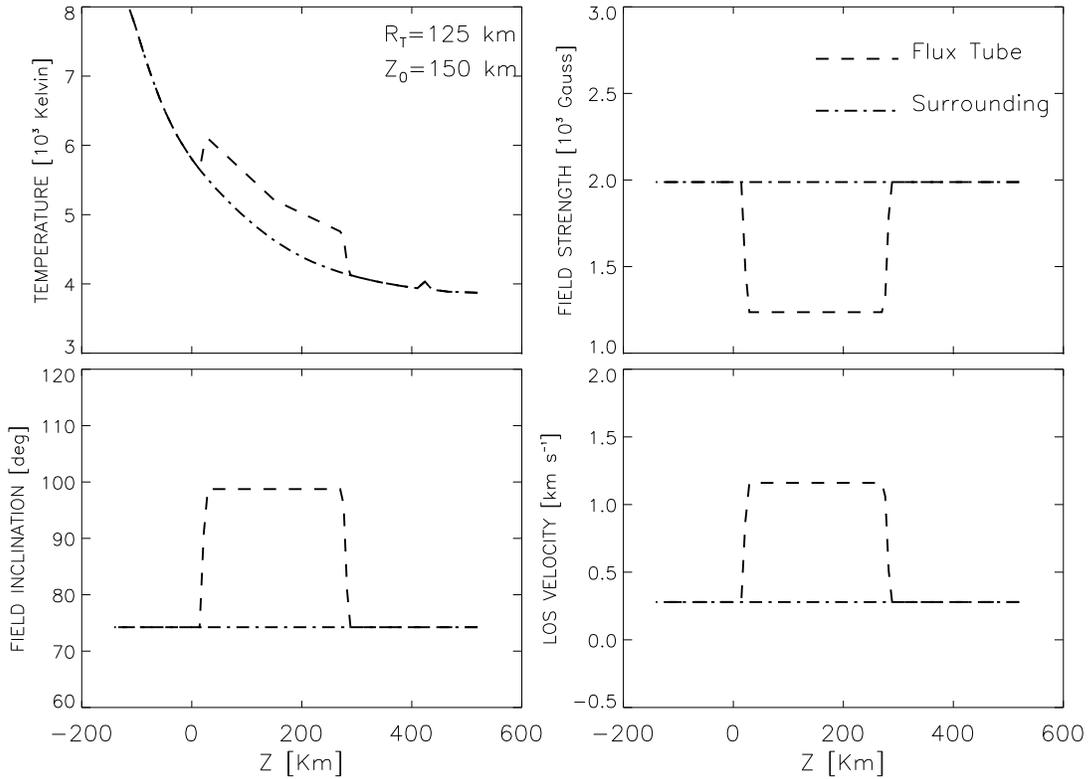}
\end{center}
\caption{Temperature (top-left panel), magnetic field strength
  (top-right panel), magnetic field inclination (bottom-left)
and line-of-sight velocity (bottom-right) for the flux tube 
atmosphere (dashed lines) and its surroundings (dot-dashed lines) 
as a function of the geometrical depth, obtained from the inversion
of the profiles in Fig.~3. The flux tube's radius returned by the inversion is
125 km and its central position is $z_{0}=150$ km.}
\end{figure*}

\subsection{General properties}%

We have taken the individual results of the inversions of the 
considered radial cuts (separately for each heliocentric angle)
at the geometrical height $z=z_{0}$ (i.e. at the location of the tube
axis), and calculated the averages at each radial position in the
sunspot. This is done individually for the flux tube component 
and the surrounding magnetic field.
The results are presented in Fig.~5 and 6 for the temperature,
line-of-sight velocity, the flux tube's filling factor $\alpha_{\rm t}$,
magnetic field zenith angle and magnetic field 
strength. In order to compare the magnetic
field inclination deduced for sunspots located at different heliocentric
positions we have converted from the observer's reference frame,
 $\gamma$, to the local reference frame. Therefore, we plot now the
zenith angle, $\zeta$ (Fig.~6; top panels). A zenith angle
smaller than, equal to or larger than 90$^{\circ}$ indicates that 
the magnetic field is inclined upwards, lies parallel to or is 
inclined downwards with respect to the solar surface. All in all, the results
for the two sets of observations are remarkably similar in spite of the
different viewing angles and the time difference of 6 days between the two
observations.

In the inner penumbra, we detect nearly, but not completely horizontal
flux tubes ($\zeta_{\rm t} \simeq 70-80^{\circ}$) that are hotter 
than their surroundings by about 500-1000 K. These flux tubes carry 
most of the Evershed flow, with LOS velocities 
in the inner penumbra ranging from
0.5 to 3 km s$^{-1}$. The magnetic field strength
in the flux tubes is around 1500 Gauss. The atmosphere surrounding
these flux tubes possesses a more vertical ($\zeta_{\rm s} \simeq
20-40^{\circ}$) and stronger magnetic field ($B \simeq$ 2300-2500).
No signatures of the Evershed flow are detected here.

As the radial distance increases, the flux tubes cool down to
temperatures similar to those of the surroundings. At large radial distances $r/R \ge
0.7$ the temperature in the flux tube component decreases even below
the surrounding temperature, although only slightly ($\simeq$ 200-300
K). At the same time the tubes become more horizontal, reaching $\zeta_{\rm t}
\simeq 90^{\circ}$, and point slightly downwards with respect to the solar
surface, $\zeta_{\rm t} \simeq 95-100^{\circ}$, near the outer edge of the
penumbra, i.e. at $r/R \ge 0.8$. In addition, their magnetic field strength
decreases slowly to 800-1000 Gauss at $r/R=0.9$ while the LOS
velocity increases monotonically in both spots, although for $\mu=0.51$
it suffers a sudden drop near the outer penumbral boundary.
The filling factor of the flux tubes (Fig.~5; bottom panels) increases
continuously from very small values at the inner penumbra $\alpha_{\rm
  t} \simeq 0.2-0.3$ until they cover almost all the resolution element
at the outer boundary $\alpha_{\rm t} \simeq 0.9$. This can be
interpreted either as an increase in the horizontal cross section of the flux
tubes or as an increase in the number of flux tubes per resolution element.

The surrounding atmosphere exhibits a quite different behaviour. 
The magnetic field strength decreases
very rapidly towards the outer penumbra, reaching similar values to
those of the flux tube's magnetic field strength ($\simeq$ 800-1000
Gauss) at $r/R \gtrsim 0.8$. The inclination of the magnetic field increases slightly with radial
distance up to $\zeta_{\rm s} \simeq 40-50^{\circ}$. The LOS
velocities remain small throughout the penumbra.

\begin{figure*}
\begin{center}
\begin{tabular}{cc}
\includegraphics[width=7.5cm]{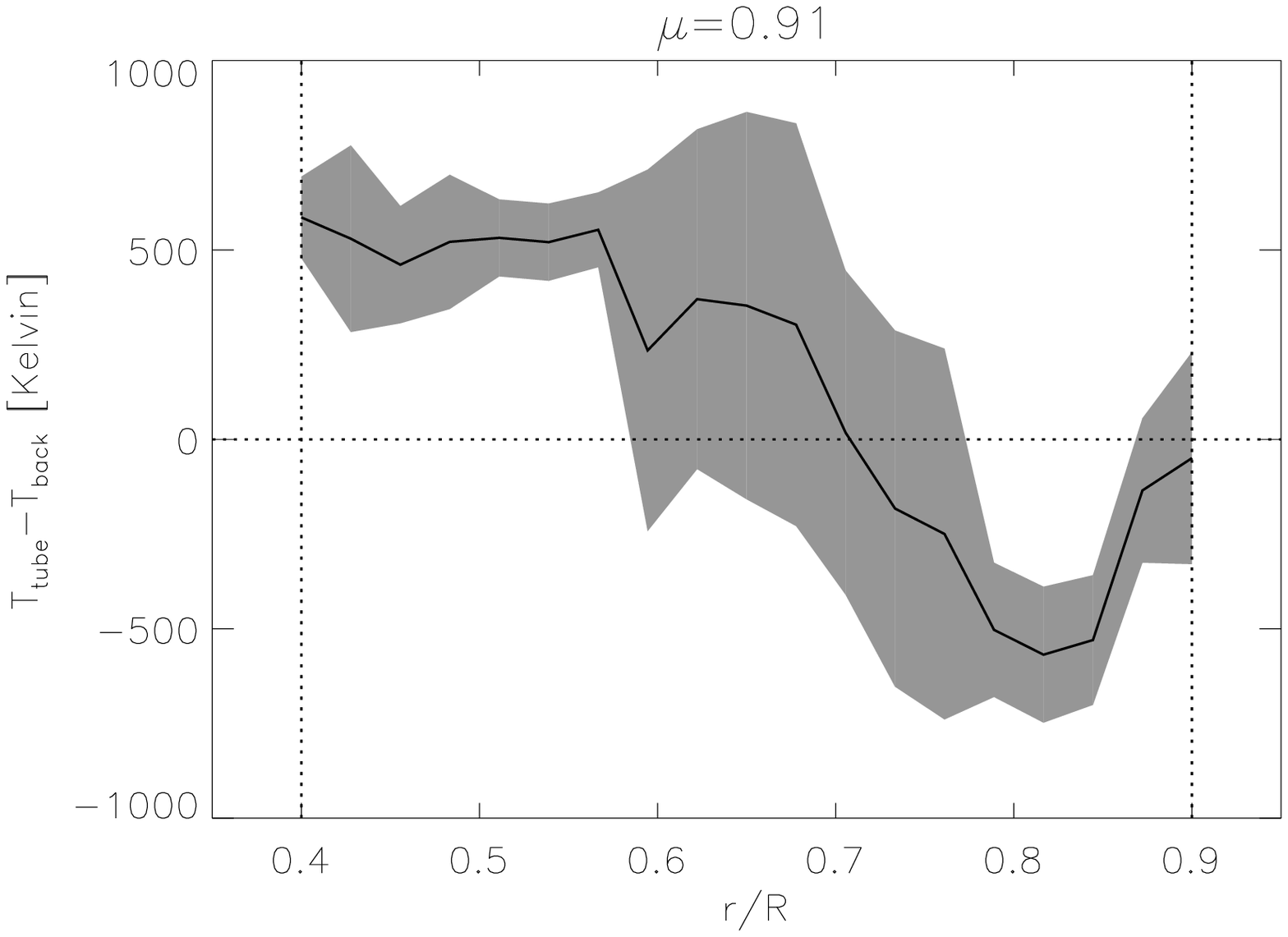} &
\includegraphics[width=7.5cm]{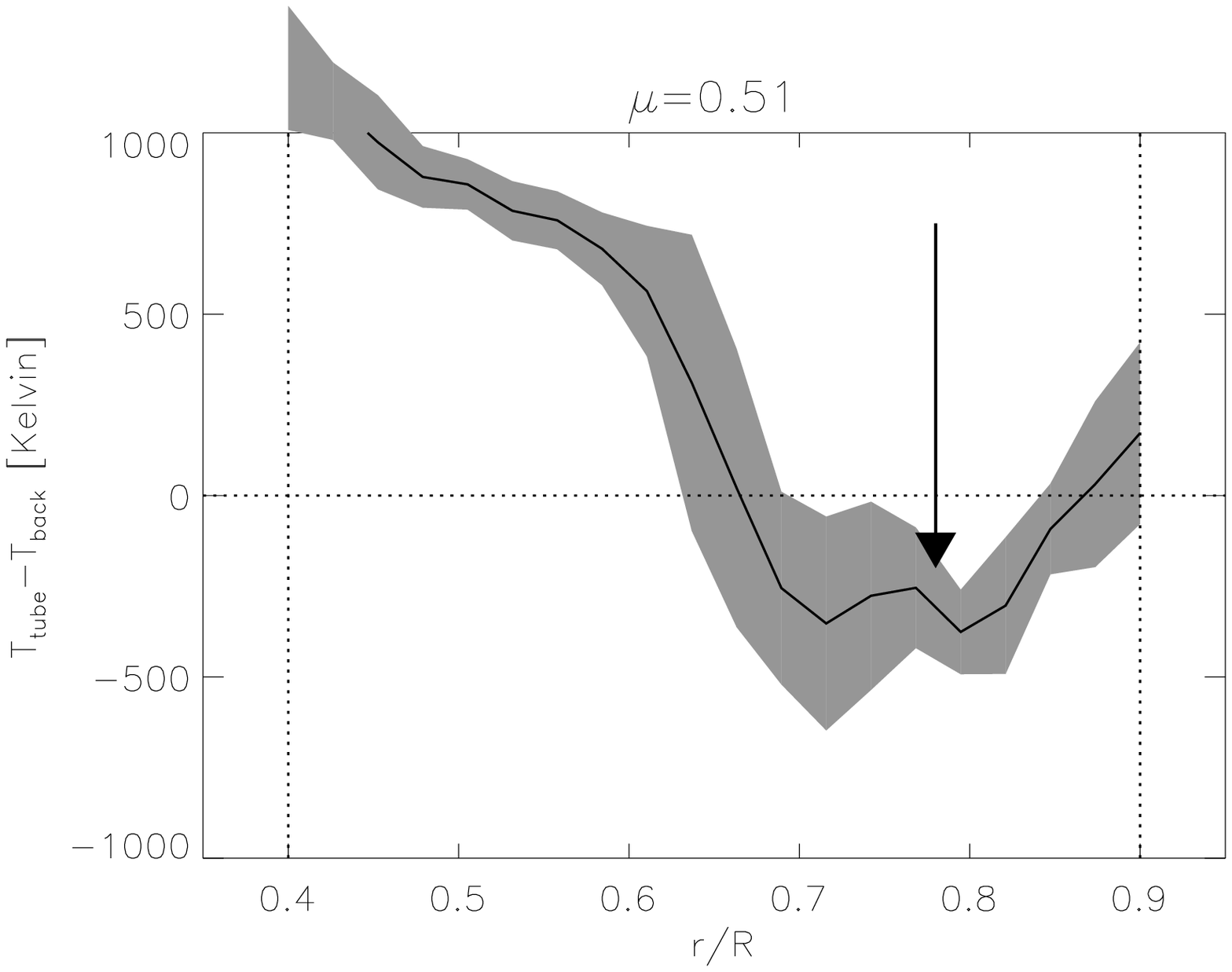} \\
\includegraphics[width=7.5cm]{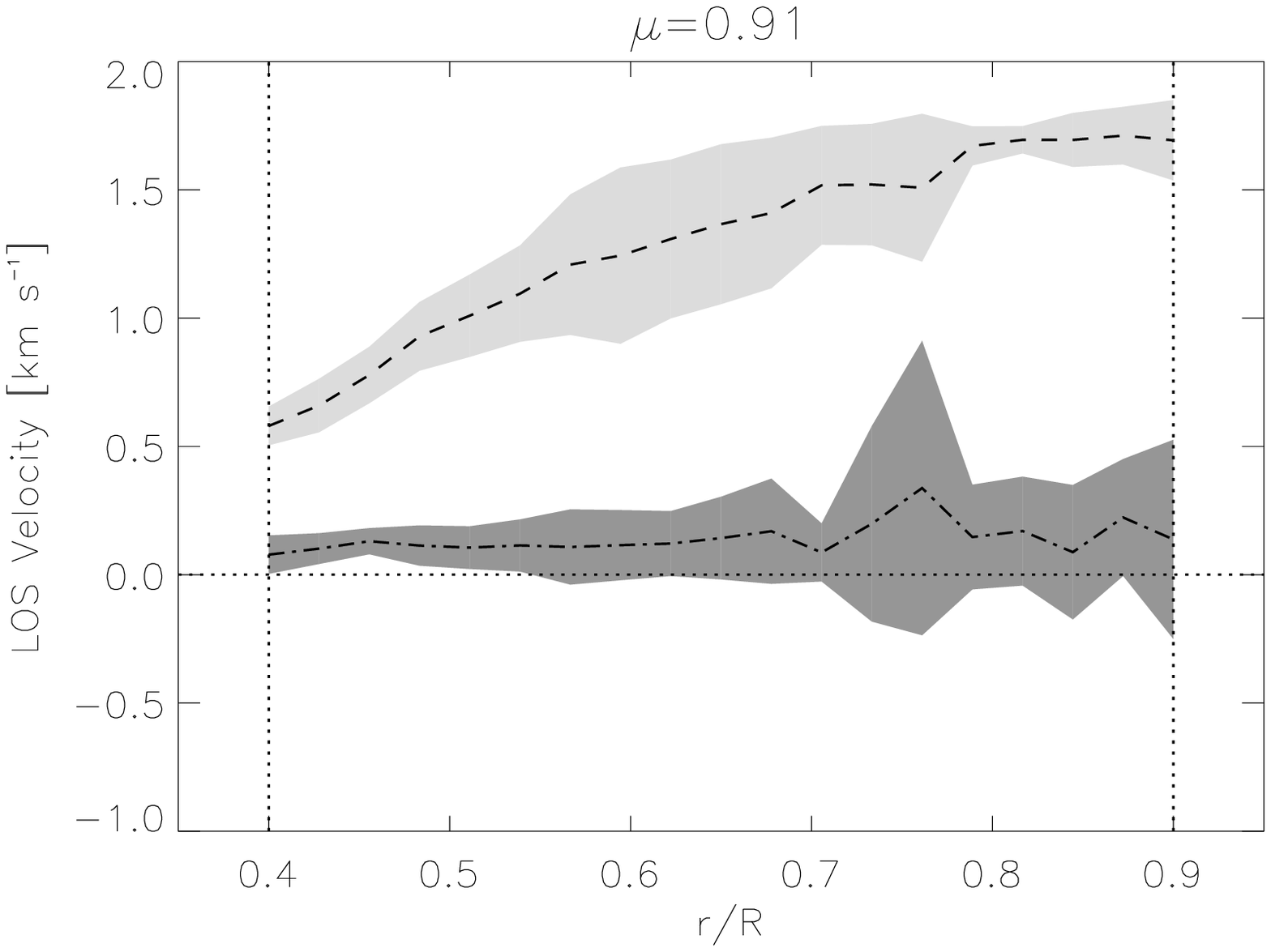} &
\includegraphics[width=7.5cm]{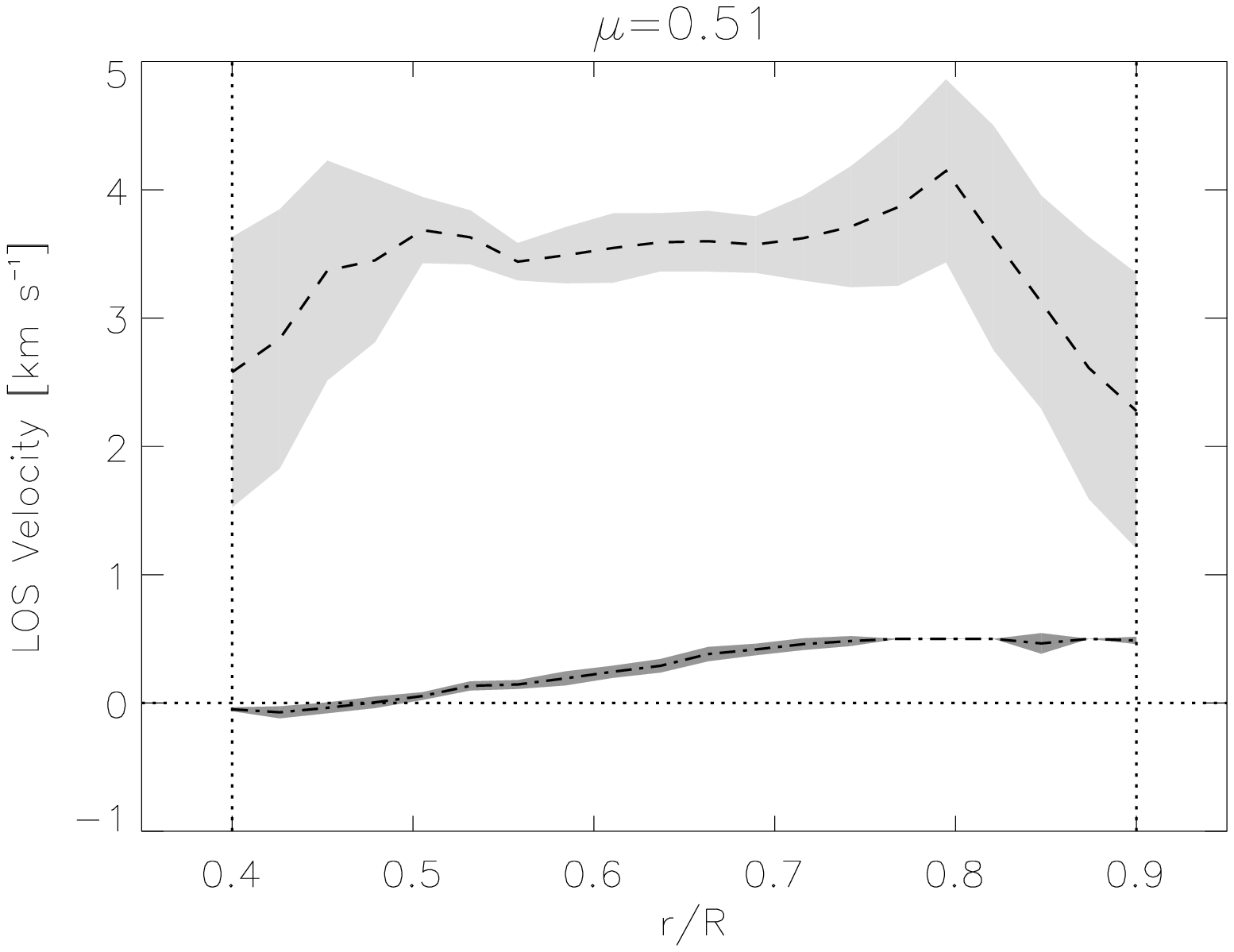} \\
\includegraphics[width=7.5cm]{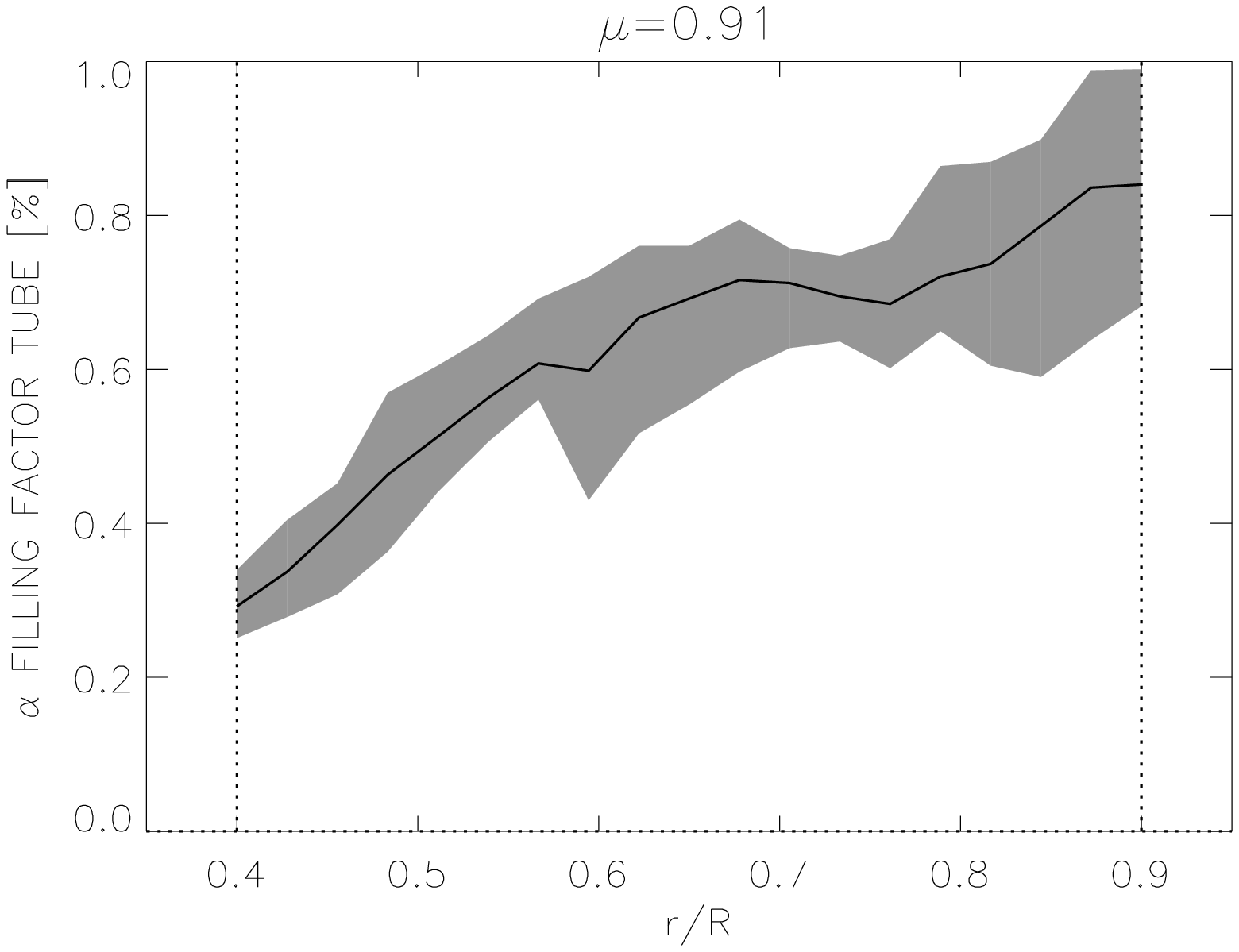} &
\includegraphics[width=7.5cm]{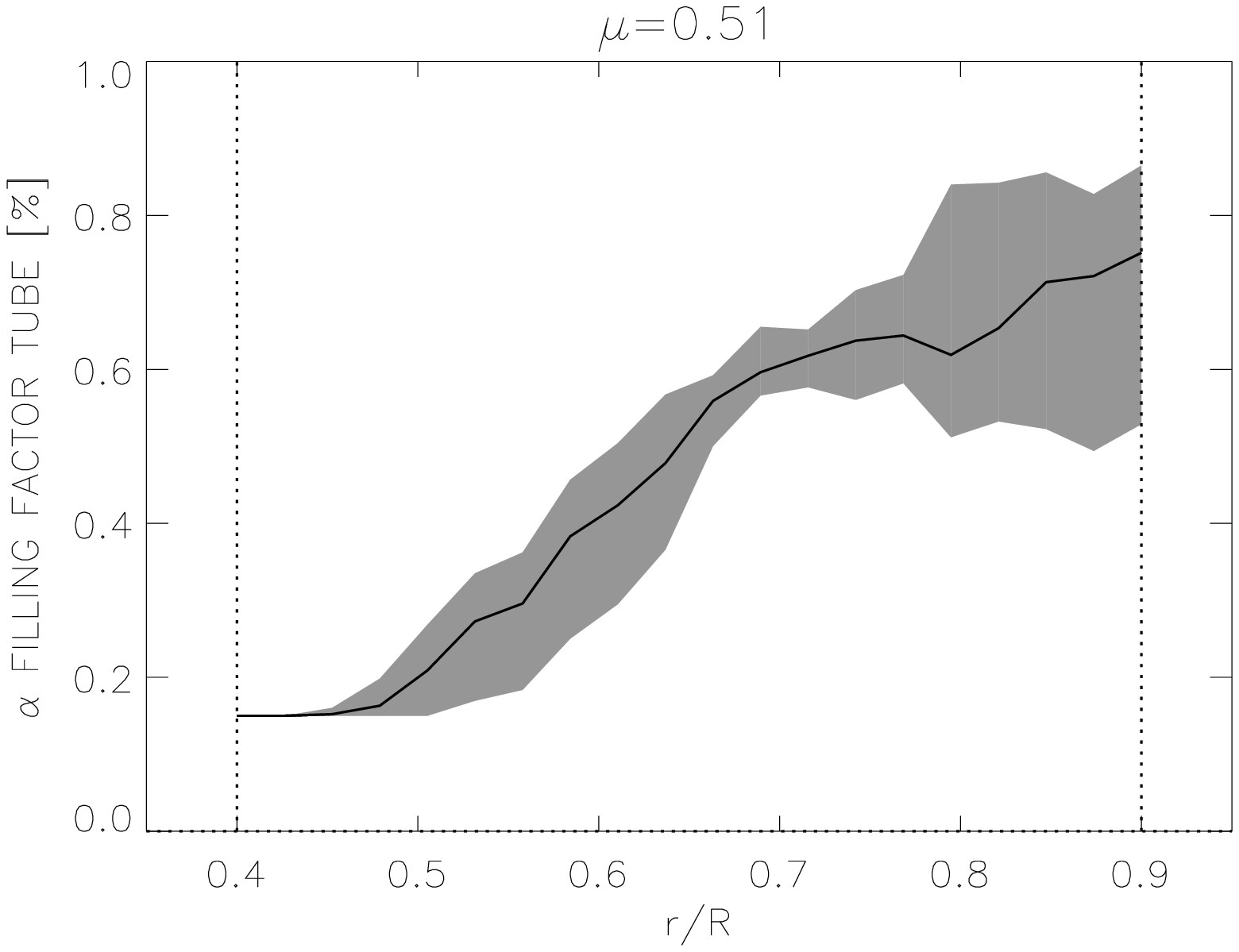}
\end{tabular}
\end{center}
\caption{Top panels: temperature difference between the flux tube
  atmosphere and its surroundings as a function of radial distance from the
  spot center. Middle panels: radial variation of the line-of-sight velocities 
  inside (dashed lines) and outside the flux tubes (dashed-dotted
  lines). Bottom panels: radial variation of the flux tube's filling
  factor, $\alpha_{\rm t}$. Left panels: sunspot at $\mu=0.91$. Right Panels:
  sunspot at $\mu=0.51$. All quantities refer to the central 
  position of the tube, $z=z_{0}$. Shaded areas denote the maximum
  deviations from the average at each radial position, obtained from 
  the individual inversions of all the
  radial cuts. The arrow indicates the approximate radial position where the
  flux tube experiences a final temperature enhancement (see Sect.~6.4).}
\end{figure*}

\begin{figure*}
\begin{center}
\begin{tabular}{cc}
\includegraphics[width=7.5cm]{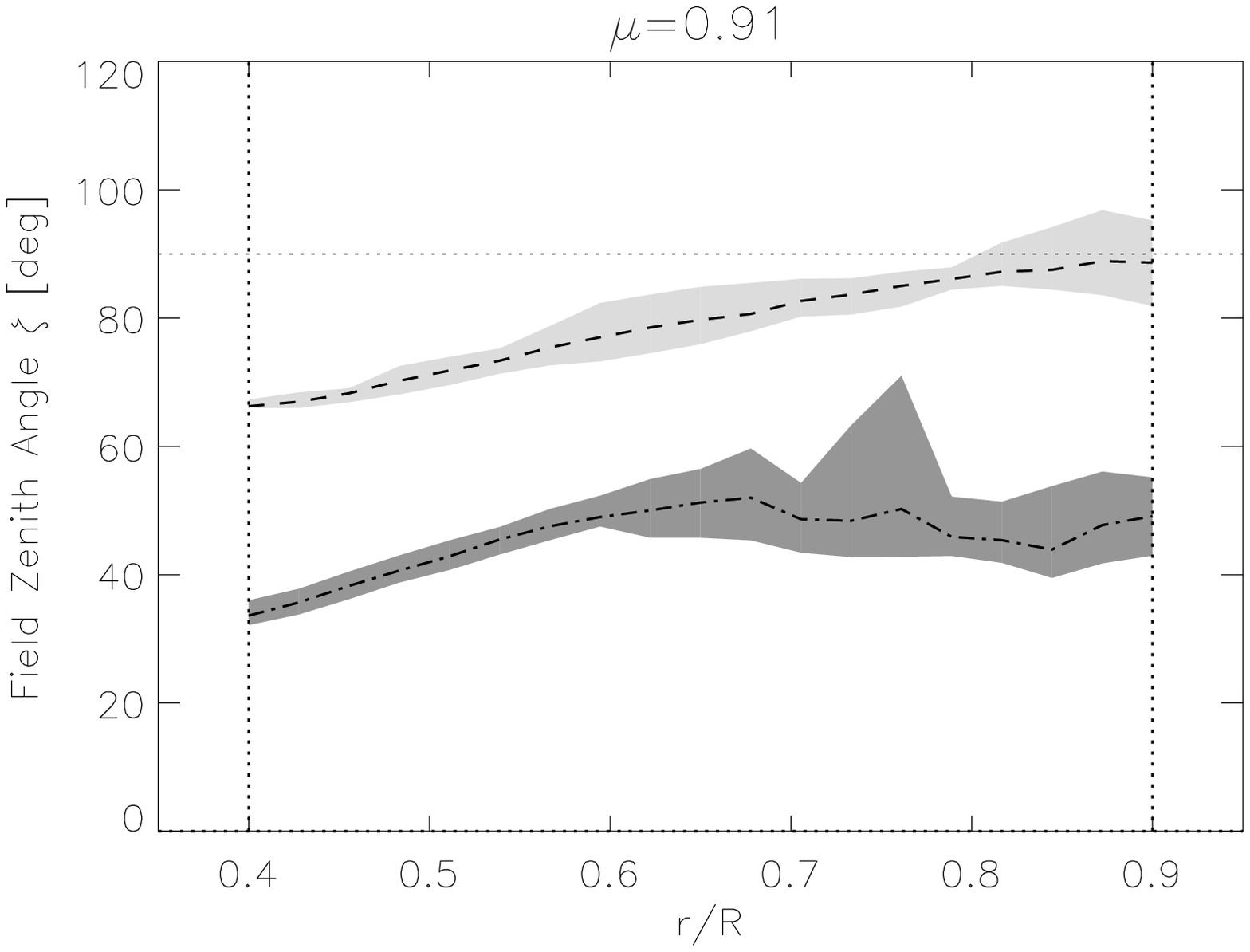} &
\includegraphics[width=7.5cm]{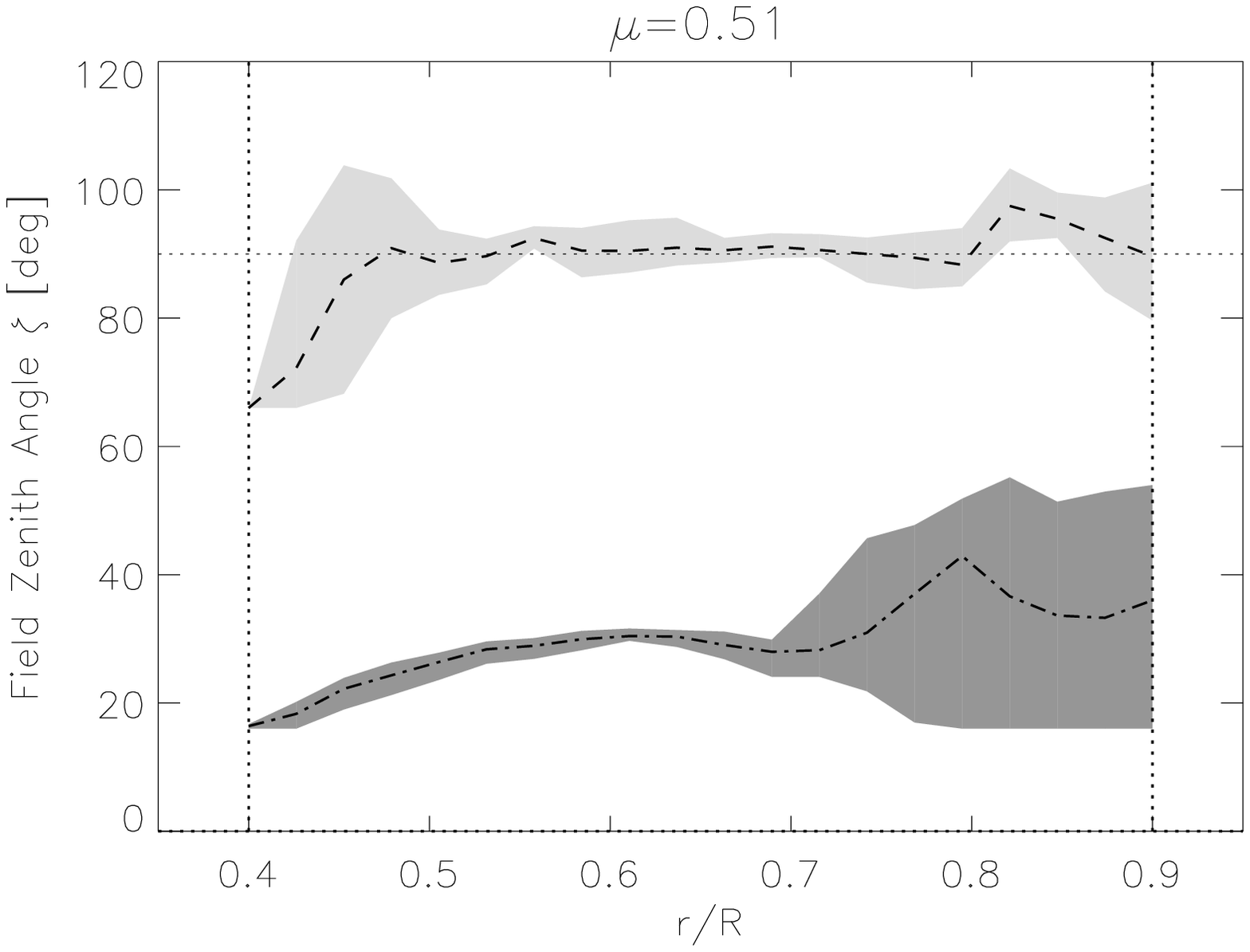} \\
\includegraphics[width=7.5cm]{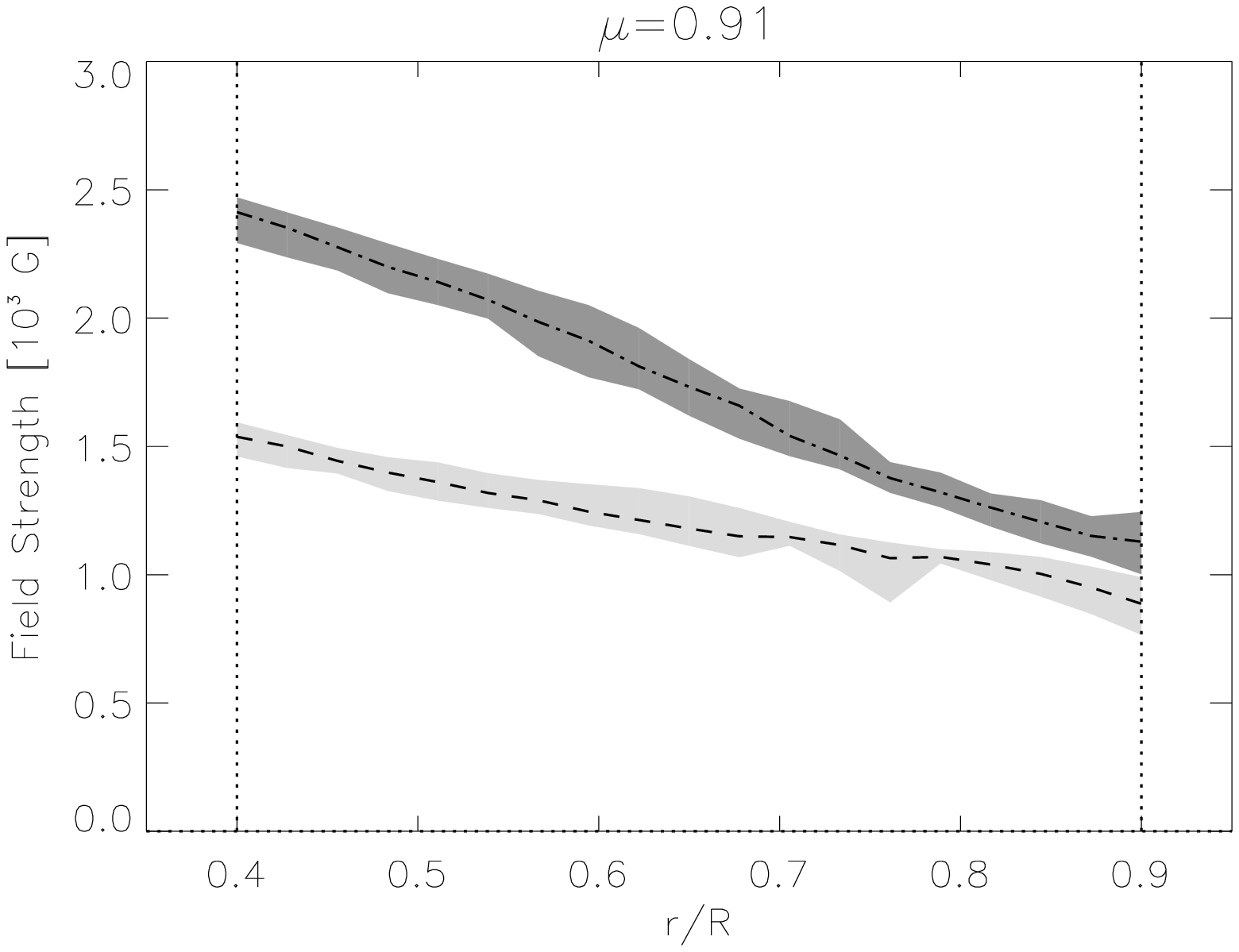} &
\includegraphics[width=7.5cm]{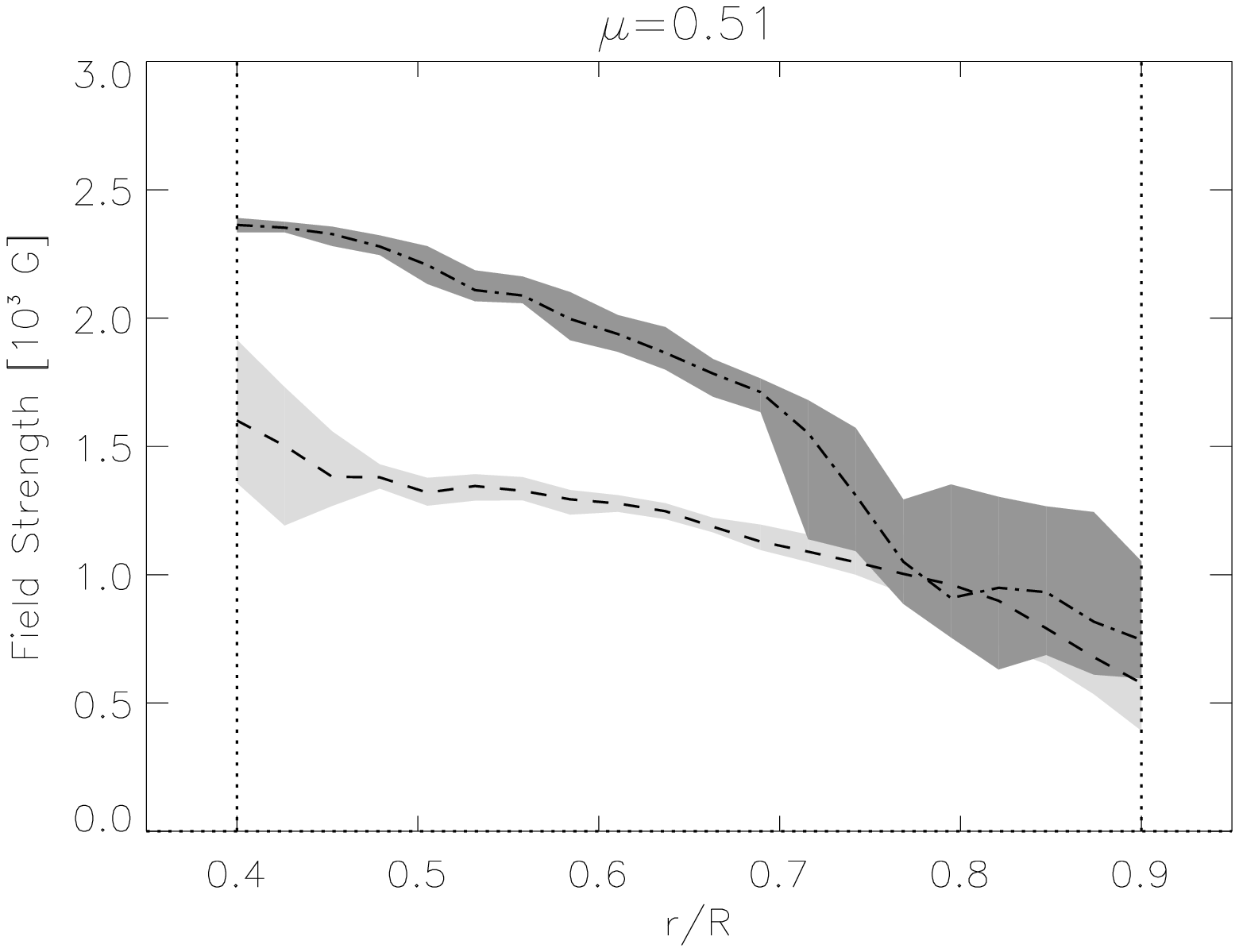}
\end{tabular}
\end{center}
\caption{The same as Fig.~5, but now for the radial variation of the magnetic
  field zenith angle (top panels) and magnetic field strength (bottom panels).}
\end{figure*}

Before discussing the individual results in more detail there are
several points which need to be clarified. First of all, we want to
stress that the results obtained from the inversions of the radial
cuts when the spot is near the disk center (Figs.~5 and 6; left panels) 
are more reliable at intermediate to large radial distances, while
results inferred from the sunspot at large heliocentric angles 
(Figs.~5 and 6; right panels) are more reliable in the 
inner penumbra. The reason for this is that the magnetic neutral 
line is located in the outer penumbral for sunspots near
the disk center, $\mu=0.91$, but it shifts towards the
umbra as the sunspot is located closer to the limb ($\mu=0.51$). 
The magnetic neutral line is where multi-lobed Stokes $V$ profiles
are commonly observed (S\'anchez Almeida \& Lites 1992; Schlichenmaier \&
Collados 2002). As already discussed in Paper I
inferred parameters from the inversion are more reliable for exactly
such complex profiles, since the signature of the two unresolved
components (flux tubes and magnetic surrounding) can be better distinguished there
(e.g. Fig.~3), allowing for a more reliable inference of their
properties. In general, however, larger and more accurate line-of-sight
velocities are obtained from the sunspot near the limb
(at all radial distances) since the Evershed flow is more aligned 
with the line-of-sight there.

We note that the possible size of the flux tubes can be much smaller 
than our spatial resolution 
of 1 arc sec (S\"utterlin 2001; S\"utterlin et al. 2004; Scharmer et al. 2002;
Rouppe van der Voort et al. 2004). Threfore the deduced properties are unlikely to correspond
to one single flux tube but rather to some average over
all the possible penumbral fibrils contained in the resolution
element. However, as one can see from Fig.~5 and 6, the inferred properties
are similar to those expected for a single flux tube
which crosses the penumbra from its inner to the outer boundary.

\section{Width of the penumbral filaments and Stokes $V$ area asymmetry}%

The area asymmetry, $\delta A$, of the circular polarization is defined as:

\begin{equation}
\delta A = \frac{\int V(\lambda) \, {\rm d} \lambda}{\int |V(\lambda)| 
\, {\rm d}  \lambda}
\end{equation}

\noindent It is different from zero whenever gradients along the line-of-sight
of the magnetic field vector and LOS velocity are present
(Landolfi \& Landi degl'Innocenti 1996). Solanki \& Montavon (1993) 
realized that the huge gradients 
needed to reproduce the area asymmetry observed in the sunspot
penumbra with the visible Fe I lines at 6301.5 and 6302.5 \AA~ (S\'anchez Almeida
\& Lites 1992) could be interpreted as a horizonal flux tube embedded in a more
vertical background. As the line of sight crosses the tube's
boundaries the physical stratifications describing the atmosphere
suffer a jump that is directly responsible for the generation of the
area asymmetry in Stokes $V$. Schlichenmaier at al. (2002) and M\"uller et
al. (2002) further investigated this issue and pointed out that the
area asymmetry observed in the visible Fe I 6301 \AA~ lines is
dominated by jumps in the magnetic field inclination, while the area
asymmetry observed in the infrared 1.56 $\mu$m lines can be explained
in terms of jumps in the magnetic field azimuth. Both cases are
compatible with the uncombed penumbra proposed by Solanki \& Montavon
(1993).

As already described in Sect.~2, in the uncombed model, the discontinuities 
needed to produce asymmetric circular polarization profiles
are located at the tube's boundaries: $z^{*}=z_0 \pm R_{\rm t}$, with
$R_{\rm t}$ and $z_0$ (the tube's radius and central position) being
free parameters of the inversion. The radius we obtain from the
inversion of all the pixels contained in the 10 radial cuts in Fig.~1 
in the same sunspot at two different heliocentric angles is,
consistently, 125 Kilometres, which is precisely the maximum value
we allow for $R_{\rm t}$\footnote{This maximum value is set in the
  inversion code so that the upper boundary of the flux tube lies below the
top of the tabulated atmosphere. This ensures that the condition 
$z(\tau_{t})=z(\tau_{s})$ for $z > z_0+R_{\rm t}$ (see Sect.~2) can be
applied.}. This means that the inversion tries to 
make the flux tube as thick as possible. The area asymmetry of the synthetic
profiles, $\delta A_{\rm FIT}$ only poorly reproduces the observed one: 
$\delta A_{\rm OBS}$ (see Fig.~7; top panels). In most cases the uncombed
model produces an area asymmetry which is very little or almost zero,
and varies over a smaller range than $\delta A_{\rm OBS}$.

The small observed area asymmetries (average of $\delta A_{\rm OBS}
\sim 3$ \%) clearly indicate that $\delta A$ is a parameter which plays a minor
role in the inversion since the shape of the circular polarization
profiles can be successfully fitted by means of profiles showing little
or even no area asymmetry at all (as the 2C model in Paper I). 
Even extremely strange profiles, such as
those presented in Fig.~3, can be fitted reasonably well. This is because
 the circular polarization profiles of the spectral lines used in this work are 
mainly affected by the presence of two different polarities in the
resolution element (background with $\gamma < 90^{\circ}$
and flux tube with $\gamma > 90^{\circ}$) rather than by discontinuities
along the line of sight in the physical parameters. This is due to the large
Zeeman splitting of these loines and the fact that they are not strongly
saturated (see Grossmann-Doerth et al. 1989). Note that Borrero et al. (2004;
paper I) also reached the same conclusion and
suggested that using the visible iron lines at Fe I 6301 \AA, 
whose area asymmetry is far more sensitive to such
discontinuities, might help to further constrain the size
of the penumbral filaments. For the case of the infrared Fe I lines 
at 1.56 $\mu$m they constrained a small region where the area
asymmetry of these lines is sensitive to gradients along the line
of sight: $\log\tau_5 \in [0,-0.5]$.

In Fig.~7 (middle panels) the position (in the
optical depth scale) of the tube's lower boundary is plotted
as a function of $r/R$. At $r/R < 0.7$ the
lower boundary is located at $\log \tau_5 > 0$. At the outer penumbra,
however, the lower boundary shifts to higher layers: $\log\tau_5 \sim
-0.25$, where the discontinuity is effective in generating area asymmetry 
(see Paper I). Note that for most of the penumbra the lower boundary also
remains below the $\log\tau_{1.56}$ level (dashed line).
For comparison we plot  in Fig.~7 (bottom panel) the observed, 
$\delta A_{\rm OBS}$ (solid line), and fitted (dashed line), $\delta A_{\rm FIT}$, 
area asymmetry for all the considered radial cuts versus $r/R$. 
$\delta A_{\rm OBS}$ increases radially in the penumbra, from
small negative values ($\sim -$2 \%) up to larger positive ones:
$\sim$ 7 \%~. Interestingly the fitted area asymmetry displays a 
similar behaviour, specially in the outer penumbra, where observed 
and fitted values become similar.

The combination of these results leads us to conclude that, in 
the inner-intermediate penumbra, where the observed area asymmetry 
in the circular polarization profiles of the Fe I 1.56 $\mu$m 
lines is small, discontinuities are not important to reproduce 
the profiles and therefore the tube radius is a parameter which is 
not well constrained from the inversion. However, in the outer
penumbra, the observed area asymmetry becomes large enough to
turn into an important ingredient to successfully fit the circular polarization
profiles. In these regions, the lower
boundary of the flux tube is located at heights where it is effective in
generating area asymmetry. As the upper boundary remains unseen
we still cannot draw any reliable conclusion on the actual size of
the penumbral filaments.

Given that Fe I lines at 1.56 $\mu$m do not see much of the tube's boundaries,
our uncombed model could in principle be simplified into a two
component model where all physical quantities are constant with height 
(e.g., making the tube's radius infinite in Fig.~4). Such
models have been previously used to study the fine structure 
of the penumbra (Bellot Rubio et al. 2003,2004; Borrero et al. 2004). 
However a feature included in the uncombed
model but has been neglected by previous two component models
remains: the use of the total pressure balance between
the flux tube and its surroundings. As we shall discuss in Sect.~6.1,
this has important consequences.

\begin{figure}
\begin{center}
\includegraphics[width=7.5cm]{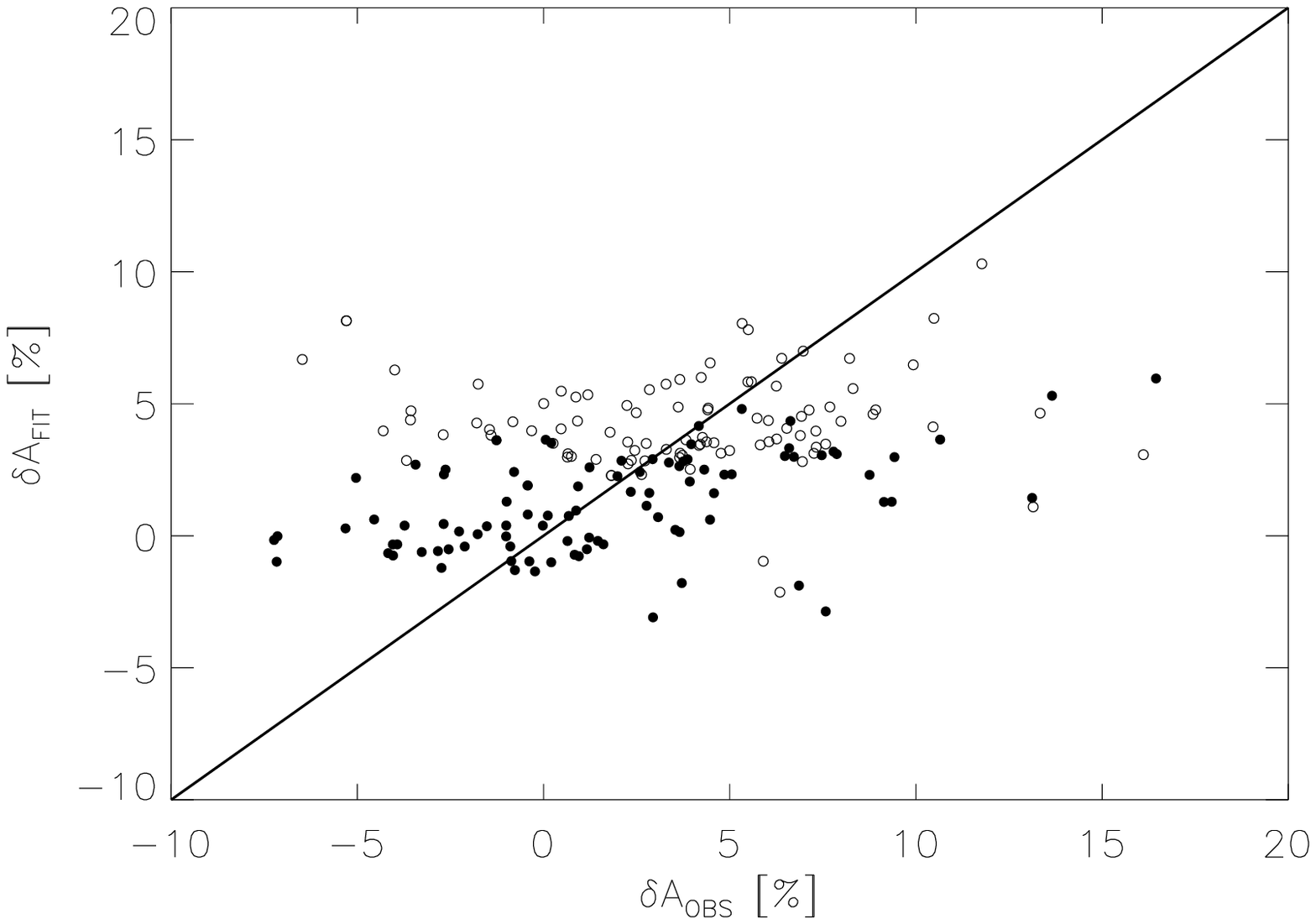} \\
\includegraphics[width=7.5cm]{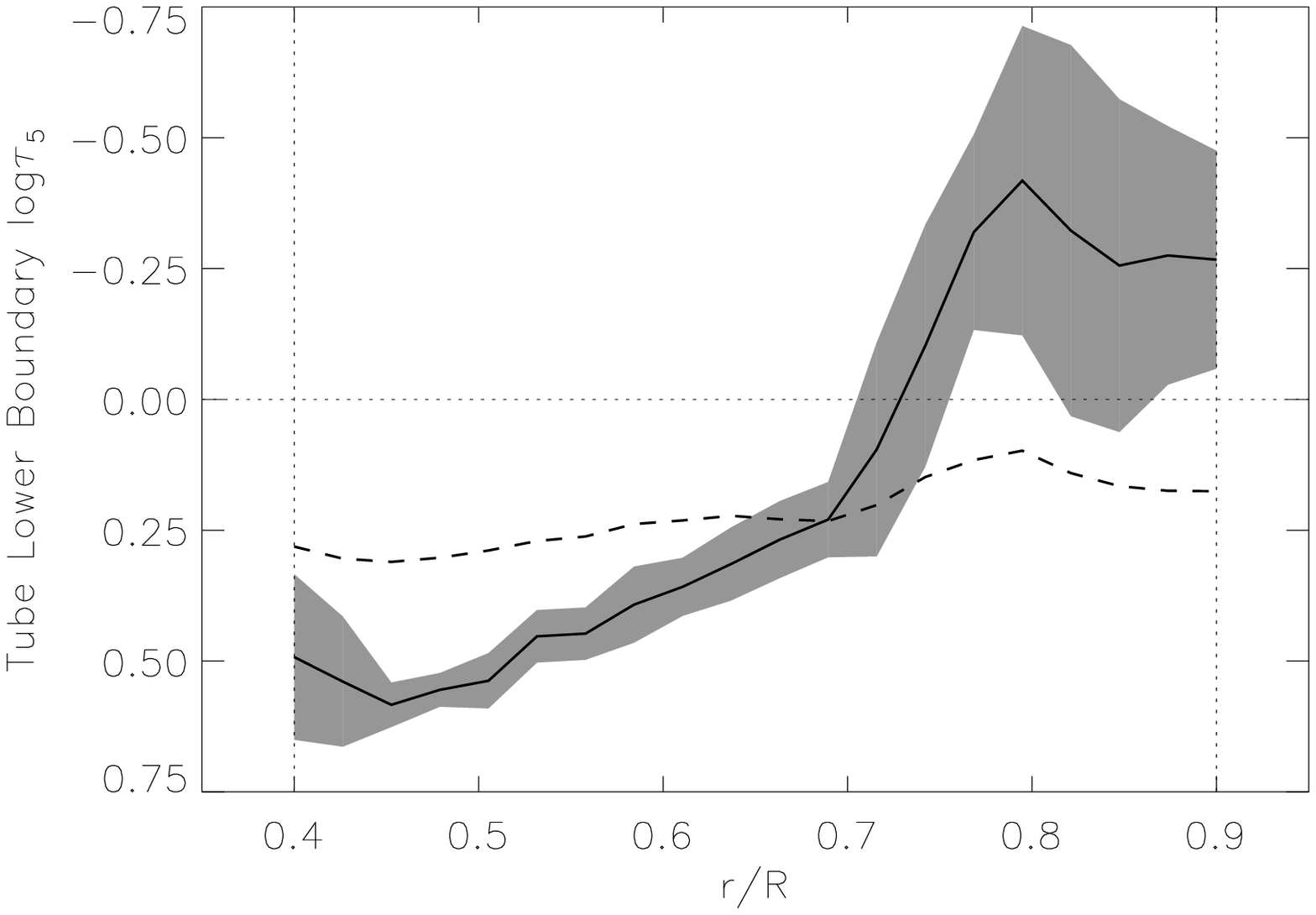} \\
\includegraphics[width=7.5cm]{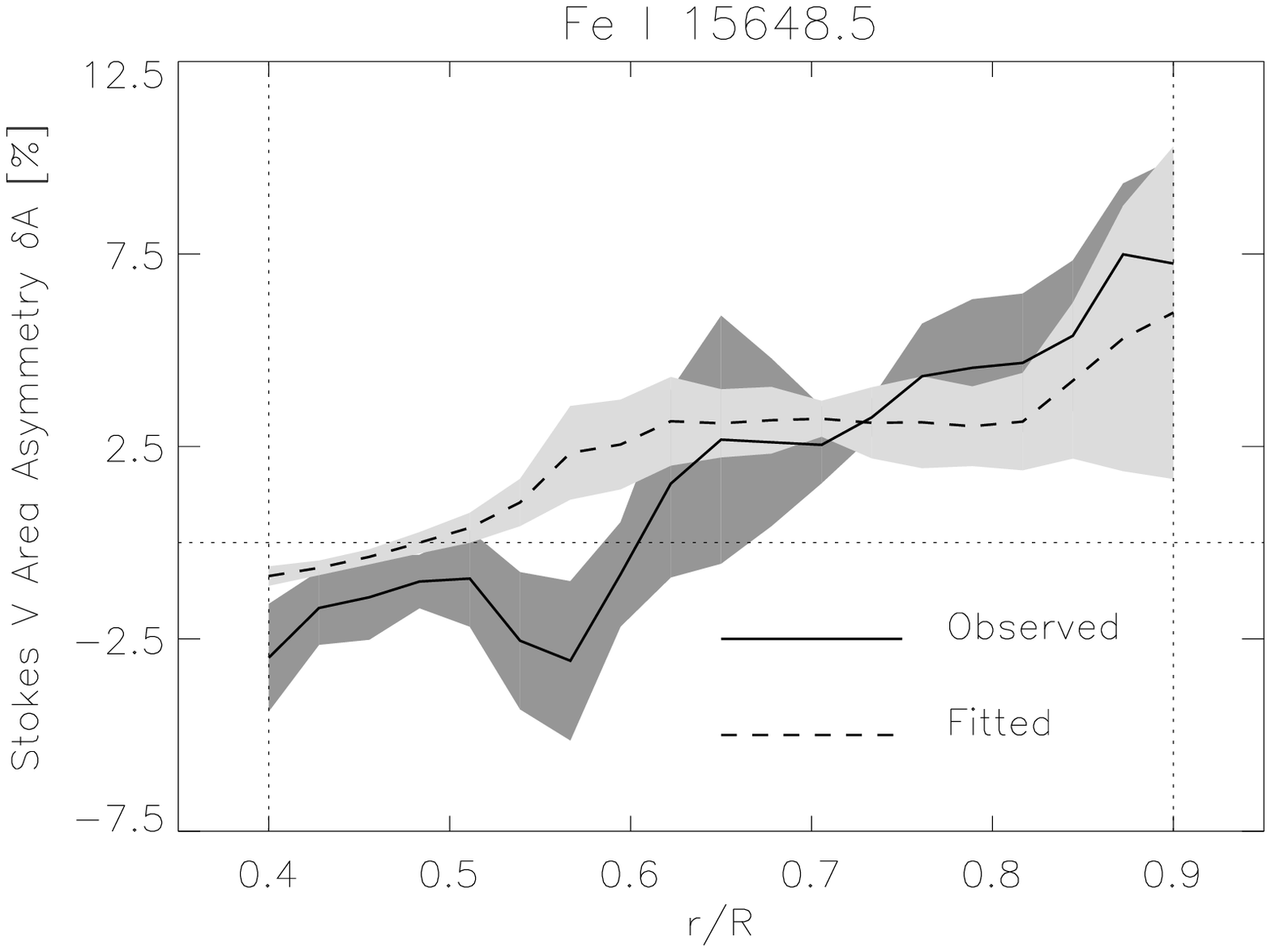} 
\end{center}
\caption{Top panel: area asymmetry from the best-fit Stokes $V$
  profiles, $\delta A_{\rm FIT}$, versus area asymmetry of the
observed Stokes $V$, $\delta A_{\rm OBS}$, for both Fe I lines for the case $\mu=0.91$. 
For $\mu=0.51$ (not shown) the result is very similar. Filled circles
  are for Fe I 15648.5 \AA, open circles for 15652.8 \AA.
Middle panel: average position, in the optical 
depth scale, of the lower boundary of the flux tube (solid
  line). The $\log\tau_{1.56}=0$ level is indicated by the dashed line.
Maximum and minimum individual deviations of the 10 radial
  cuts considered are indicated by the shaded area.
Bottom panel: radial variation of the
fitted (dashed line) and observed area asymmetry (solid
  line) of Fe I 15648.5 \AA~.}
\end{figure}

\section{Nature of the Evershed Flow}%

Previous observational analyses, which neglected the fact that
in our resolution element we have mixed signals coming from the flux
tubes and their magnetic surroundings, obtained that the 
magnetic field strength showed
a strong radial decrease from $B_{\rm inner} \simeq 2500$ G to $B_{\rm
  outer} \simeq 1000$ G (Beckers \& Schr\"oter 1969; Wittmann 1974; Lites
\& Skumanich 1990; McPherson et al. 1992; Lites et al. 1993; 
Keppens \& Mart{\'\i}nez Pillet 1996; Stanchfield et al. 1997). 
This implies that the magnetic field is larger at the inner footpoint
of a siphon flow carrying loop than in the outer one, making it difficult
for this mechanism to work.

If the fine structure of the penumbra is taken
into account (R\"uedi et al. 1998, 1999, Borrero et al. 2004) a more 
favourable situation appears: an inclined magnetic field whose
strength rapidly decreases with radial distance and an almost
horizontal magnetic field carrying the Evershed flow with a far 
smaller radial drop in field strength are inferred (see Fig.~6; bottom panels).
The small decrease in the field strength for the component carrying
the Evershed flow still implies a stronger magnetic field at small $r/R$, but
now the difference between inner and outer penumbra is significantly reduced to $\simeq 300-500$ G. 
Although this finding on its own is insufficient to produce an outwards
accelerated flow it represents a more favourable situation for the siphon flow
mechanism, specially if we take into account that, as noticed by Montesinos \&
Thomas (1997), the inner and outer footpoints, as measured by
observations, are not necessarily at exactly the same height.

\subsection{Gas pressure gradient}%

According to Fig.~6 (bottom panels) the magnetic field strength of 
the flux tubes is much smaller than that of the surrounding atmosphere 
in the inner penumbra at a geometrical height $z=z_0$ (central position of the
tube). By requiring total pressure balance between the flux tube
interior and the external atmosphere (Eq.~4) one can easily deduce
that $P_{\rm gas,t}(z_0) \gg P_{\rm gas,s}(z_0)$. At large
radial distances the situation is such that the magnetic field
strength in the flux tube and its surrounding atmosphere becomes
very similar and therefore: 
$P_{\rm gas,t}(z_0) \simeq P_{\rm gas,s}(z_0)$.
In Fig.~8 we plot the actual $\Delta P_{\rm gas}=
P_{\rm gas,t}(z_0)-P_{\rm gas,s}(z_0)$ as a function
of radial distance in the penumbra (top panel for $\mu=0.91$
and bottom panel for $\mu=0.51$). As can be seen that this difference
decreases almost linearly with $r/R$, reaching values close to zero near the outer
penumbral boundary.

Unfortunately, this effect does not imply a radial decrease in the 
gas pressure inside the flux tube, because of the unknown 
radial behaviour of the external pressure 
$P_{\rm gas,s}(z_0)$. In particular, its calculation is
ill-posed by the boundary condition described in Sect.~2, where
a value of the gas pressure at the highest photospheric layers
is prescribed for the surrounding atmosphere.
This is done for all pixels independently of their position on the spot:
$P_{\rm gas,s}(\tau_{\rm max})=P_0$. Consequently,
the geometrical height scale $z_{\rm s}(\tau_{\rm s})$ in the
surrounding atmosphere does not take into account pixel to pixel
variations of the absolute geometrical height scale (i.e. Wilson
depression). The radial variation of
the difference in gas pressure between
the flux tube and the surrounding atmosphere, however,
only depends on the difference in the magnetic field strength 
between them and is calculated under the condition of total
pressure balance (magnetohydrostatic constraint). Although it cannot 
directly prove that there is a radial decrease in the gas pressure 
along the flux tube axis, it does provide a strong indication that
this is indeed the case.

\begin{figure}
\begin{center}
\includegraphics[width=7.5cm]{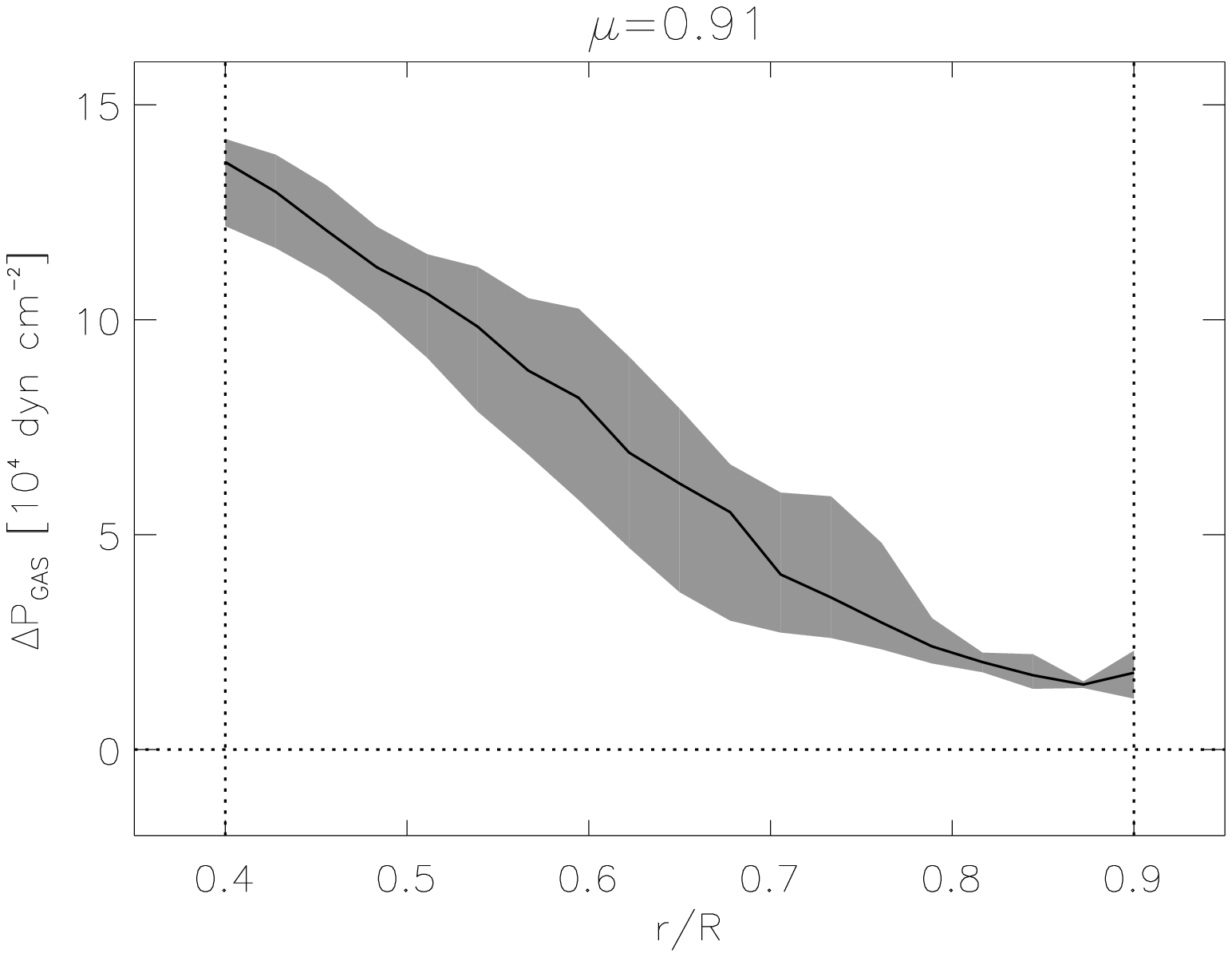} \\
\includegraphics[width=7.5cm]{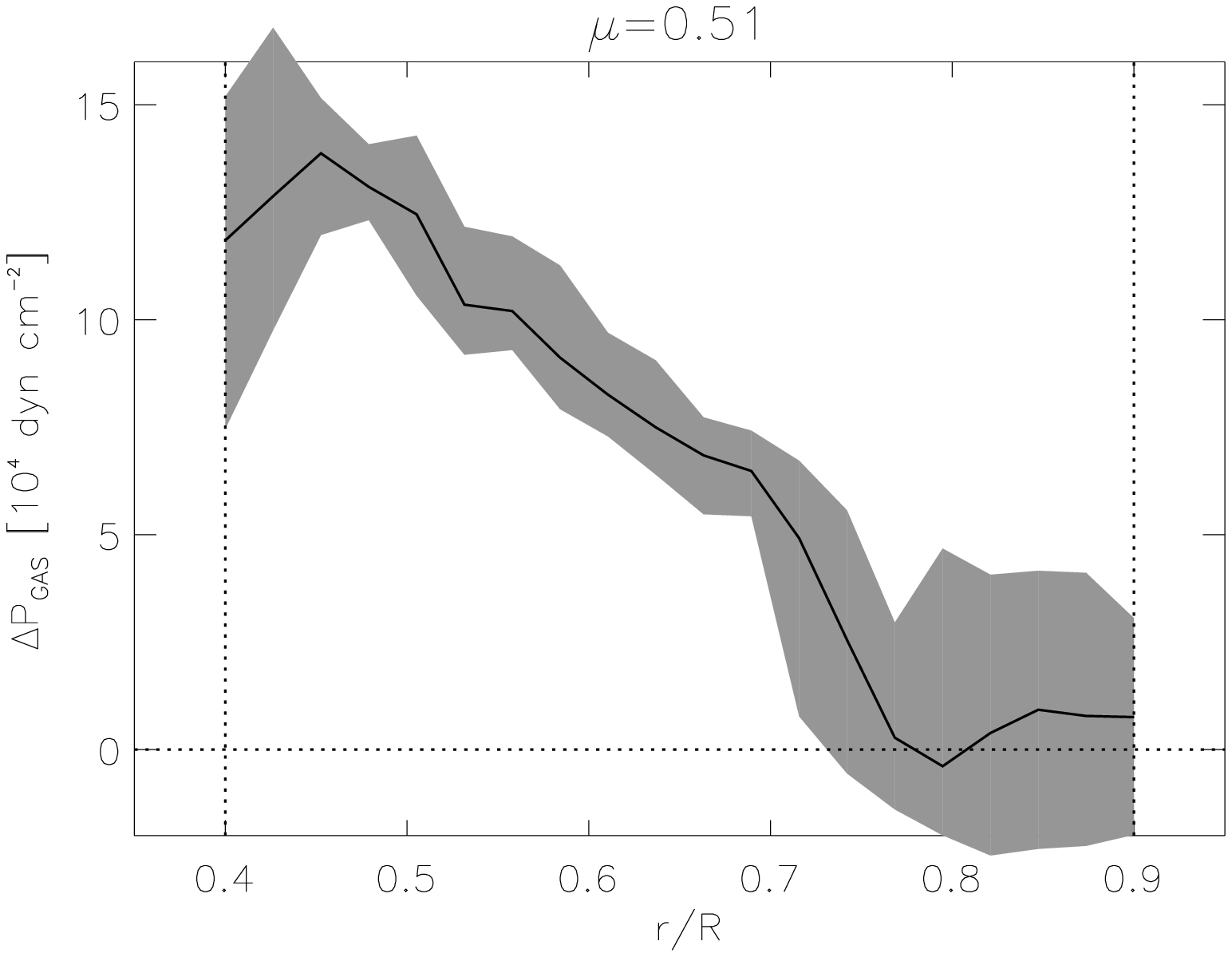}
\end{center}
\caption{Radial variation of the gas pressure difference, $\Delta
  P_{\rm gas}$, between the flux tube and its surrounding 
  atmosphere at $z=z_0$. Top panel: for NOAA 8706 at an
heliocentric angle of $\mu=0.91$. Bottom panel: the same for
$\mu=$0.51.}
\end{figure}

\subsection{Cooling flux tubes}%

Our results show that in the inner penumbra the
flux tubes are hotter (by roughly 500-1000 K)
than their surroundings. As we move to larger radial distances
the temperature becomes very similar inside and outside the flux tube.
This is in very good agreement with the theoretical predictions
of the moving flux tube model (Schlichenmaier et al. 1998a, 1998b). 
The result that at a given point, $r/R \simeq 0.7$, the flux tube becomes cooler
than its surroundings can be explained by assuming that
the flow becomes adiabatic due to a less efficient radiative exchange 
in the higher parts of the anchored flux tube (Montesinos \& Thomas
1997). The temperature increase seen at larger radial distances is explained in
terms of a shock front (see Sect.~6.3).

This picture indicates that at the umbral-penumbral boundary, a first
set of hot flux tubes (small ones or a small number of them
according to the tiny filling factor inferred; see Fig.~5 bottom panels) 
, appear. Remarkably, these flux tubes
that have just emerged already carry strong velocities (see Fig.~5 middle panels), 
perhaps as an indication that most part of the plasma 
acceleration has already occurred.
Again this is in very good agreement with the moving flux tube simulations
(see Schlichenmaier et al. 1998a,1998b) where the acceleration
takes place in the inner penumbra, where hot flux tubes carrying
plasma at about 8000-10000 K rapidly cool down. The opacity in
such extremely hot tubes would be too large to make them 
observable; therefore what we see is perhaps, the latest stages of
this cooling process. Needless to say, hotter flux tubes in the 
inner penumbra were already expected considering that
at small radial distances, the magnetic field strength in the flux tubes
is much smaller than in the surrounding atmosphere, and therefore, in order to keep
the horizontal total pressure balance between these two atmospheres the gas
pressure in the flux tube component has to increase.

\subsection{Shock front}%

Another point of special interest is the magnitude of the
Evershed flow. Theoretical models predict different values
for the speed of the flow inside penumbral flux tubes. Stationary siphon
flow models (Thomas \& Montesinos 1991,1993; Montesinos \& Thomas 1997)
distinguish between subcritical and supercritical velocities
with respect to the characteristic critical tube speed 
(Thomas 1988, Ferriz Mas 1988). Transitions between both regimes can be present within the
flux tube, leading to shock fronts. Time-dependent simulations
of thin flux tubes in the penumbra predict supercritical flows in most of the
penumbra regardless of whether the flux tubes remain horizontal beyond the
visible limit of the sunspot (Schlichenmaier et al. 1998a,1998b) 
or bend back within the penumbra (Schlichenmaier 2002).
Taking into account the fine structure of
the penumbra, large velocities ($> 4$ km s$^{-1}$) are 
favoured (Wiehr 1995, del Toro Iniesta et al. 2001; 
Bellot Rubio et al. 2003; Bellot Rubio et al. 2004).
Most of these values have been obtained under the
assumption that the magnetic field and the velocity vectors
are mutually parallel (see Bellot Rubio et al. 2003).

The absolute
flux tube velocity is: $v_{\rm t}=v_{\rm los,t}/\cos\gamma_{\rm t}$, 
where $\gamma_{\rm t}$ is the inclination of the tube's magnetic field 
vector with respect to the observer (both $v_{\rm los,t}$ and $\gamma_{\rm t}$
are obtained from the inversion). $v_{\rm t}$ is plotted in
Fig.~9 (top left panel) for the sunspot at an heliocentric angle
of $\mu=0.51$\footnote{Results for $\mu=0.91$ are not considered, since
  dividing by the cosine of inclination angles close to 90$^{\circ}$ enhances any error in
line-of-sight velocities}. Also plotted are the local sound speed 
$c_{\rm s} \sim T^{1/2}$ and the tube's critical speed
$c_{\rm t}=c_{\rm s} c_{\rm a}/\sqrt{c_{\rm s}^2+c_{\rm a}^2}$, 
with $c_{\rm a} \sim B \rho^{-1/2}$ being the Alfv\'en speed. 
The velocity in the flux tube always remains subsonic, although 
we cannot rule out the possibility for supersonic values 
in the penumbra (del Toro Iniesta et al. 2001; Penn et al. 2003) 
given the limited number of radial cuts we are considering.

At almost all radial distances, $v_{\rm t} < c_{\rm t}$ as well,
 except for a few regions in the inner ($r/R \simeq 0.4$), and the outer
penumbra ($r/R \simeq 0.78$). Given the error bars, we do not deem
the first case to be reliable. However, for large radial distances (arrow
in Fig.~9; top left panel) it seems plausible that the velocity becomes supercritical.
Remarkably, after this happens, the velocity suddenly drops again by roughly 2 km s$^{-1}$
to subcritical values at larger distances: $r/R > 0.85$. 
This is accompanied by an increase in the 
flux tube temperature of about 300-400 K at these locations (see arrow 
in Fig.~5; top right panel).
Indeed, this is the behaviour expected from a transition 
between supercritical to subcritical velocities in penumbral flux
tubes (Montesinos \& Thomas 1997) produced by a shock front that 
dissipates kinetic energy and heats the gas. The presence of shocks
in the siphon flows has been predicted by numerous authors
(Meyer \& Schmidt 1968; Degenhardt 1989,1991; Thomas \& Montesinos
1991, 1993; Montesinos \& Thomas 1997; Schlichenmaier et al. 1998a, 1998b;
 Schlichenmaier 2002), but this is the first time that direct observational 
evidence of such a shock front within the penumbra is provided.

Shock fronts, if present, are likely to produce an enhancement
both in the line width (e.g. FWHM) and in the equivalent width (see
Degenhardt et al. 1993; Solanki et al. 1996). In the penumbra, the
radial variation of the line width is dominated by the
Zeeman splitting (magnetic field). However, the magnetic field
affects the equivalent widths of the comparatively weak
(i.e. unsaturated) lines considered here only very slightly.
In Fig.~9 (top right panels) we plot the radial variation of the equivalent 
width for Fe I 15648.5 \AA.
It clearly shows an enhancement ($\simeq 10$\%) at 
roughly the same radial distance where the flow speed becomes
supercritical (see vertical arrow). Three different radial positions, 
corresponding to locations before, during and after the shock have
been marked with open circles. Intensity profiles for Fe I 15648.5 \AA~
at those locations have been extracted and plotted together in Fig.~9
(bottom left panel). The intensity profile before the shock (dashed line) shows
an enhancement in the red wing which is produced by the strongly red
shifted flux tube contribution (i.e. satellite line; see Stellmacher \& Wiehr
1980; Wiehr et al. 1986; Wiehr 1995,1997).
During the shock (solid line) the equivalent width of the redshifted
component is enhanced. This broadening is likely to be produced 
by a new structure in our resolution element that our model 
does not account for. Therefore, the
only way for the inversion code to make the profile broader 
(within the constraints of the chosen model) is to increase
the micro and macroturbulent velocities. In Fig.~9
(bottom right panel) the line broadening velocity in the flux tube, defined as:
$v_{\rm broad}=\sqrt{v_{\rm mic,t}^2+v_{\rm mac,t}^2}$ is plotted.
It shows a peak exactly at the radial positions at which the flow speed
becomes supercritical. This provides a strong indication that this
spectral signature really corresponds to a shock front in the flux
tube at large radial distances in the penumbra.

In our observations the shock seems to occupy 1.2-1.4 arc sec. The
fact that we do not observe a jump over a smaller radial range is
likely to be caused by smearing effects introduced by seeing.
Finally, it is important to recall that this result 
is based on some assumptions that must be considered
carefully. In particular the density inside the flux tube, 
$\rho_{\rm t}$, which is obtained through the gas pressure and temperature by
applying the ideal gas equation (see Sect. ~2), does not satisfy
vertical hydrostatic equilibrium. The Alfv\'en speed, and
therefore the tube's critical speed, are affected by any inaccuracies in the
density, which introduces some uncertainties.

\begin{figure*}
\begin{center}
\begin{tabular}{cc}
\includegraphics[width=7.5cm]{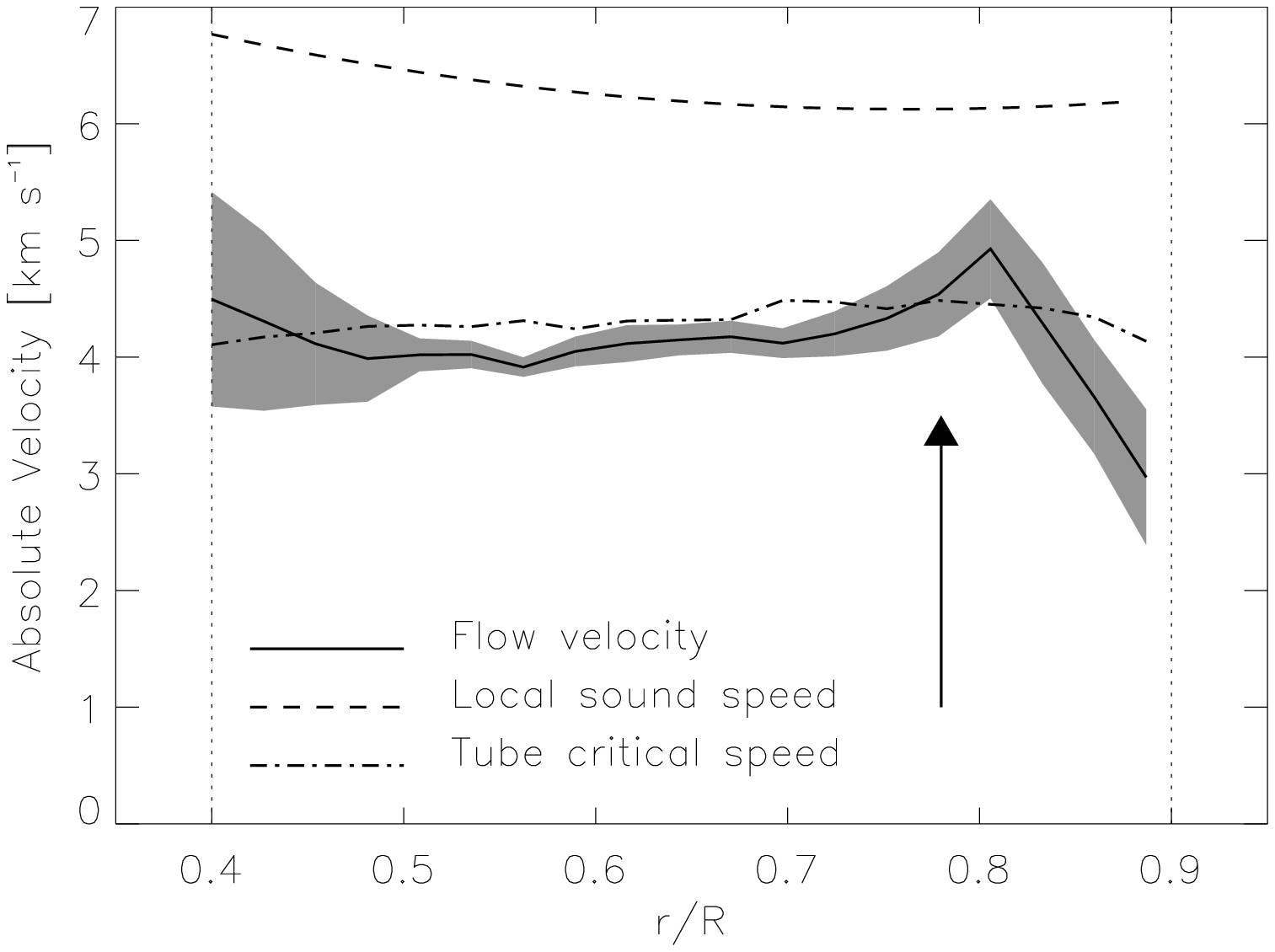} &
\includegraphics[width=7.5cm]{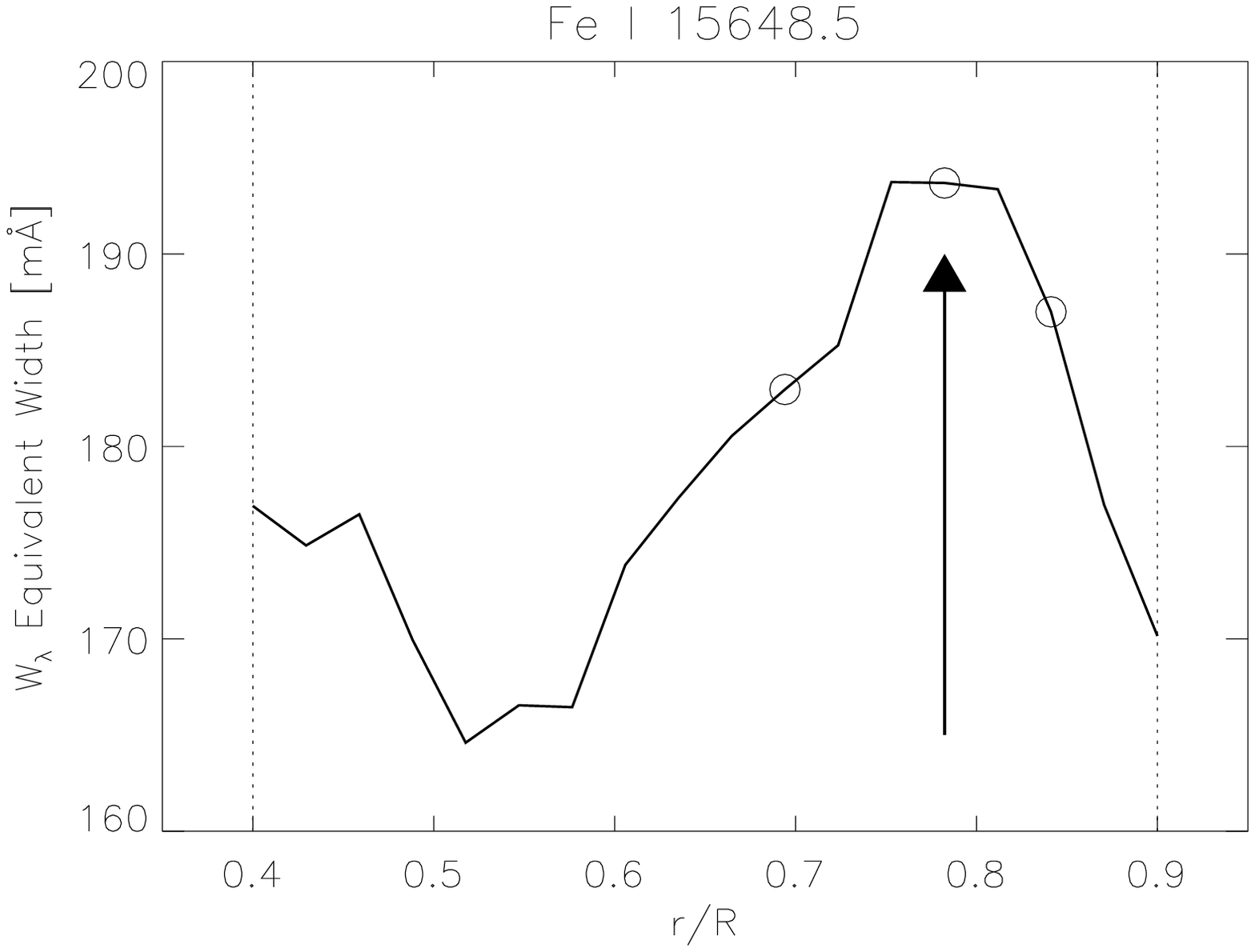} \\
\includegraphics[width=7.5cm]{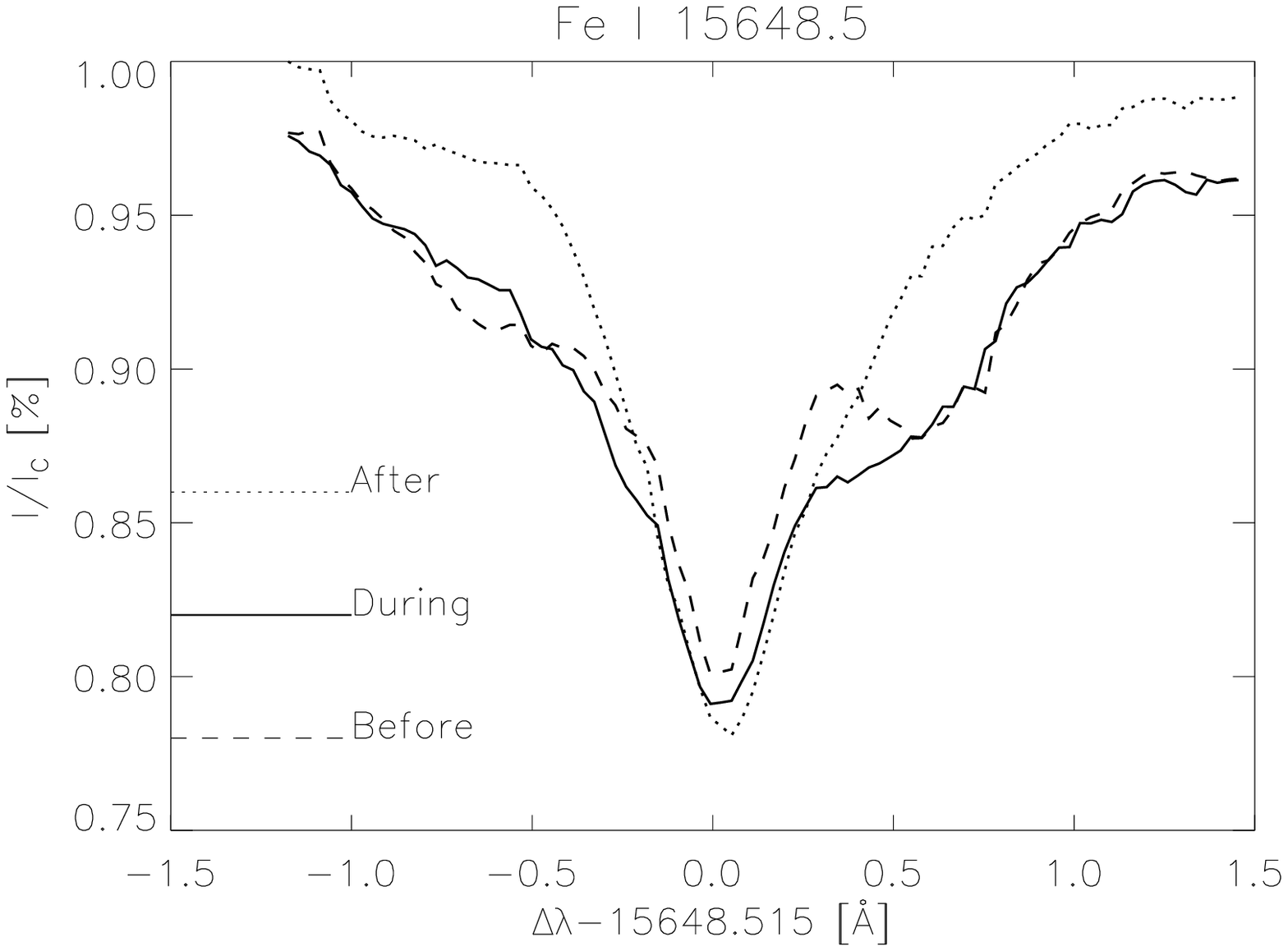} &
\includegraphics[width=7.5cm]{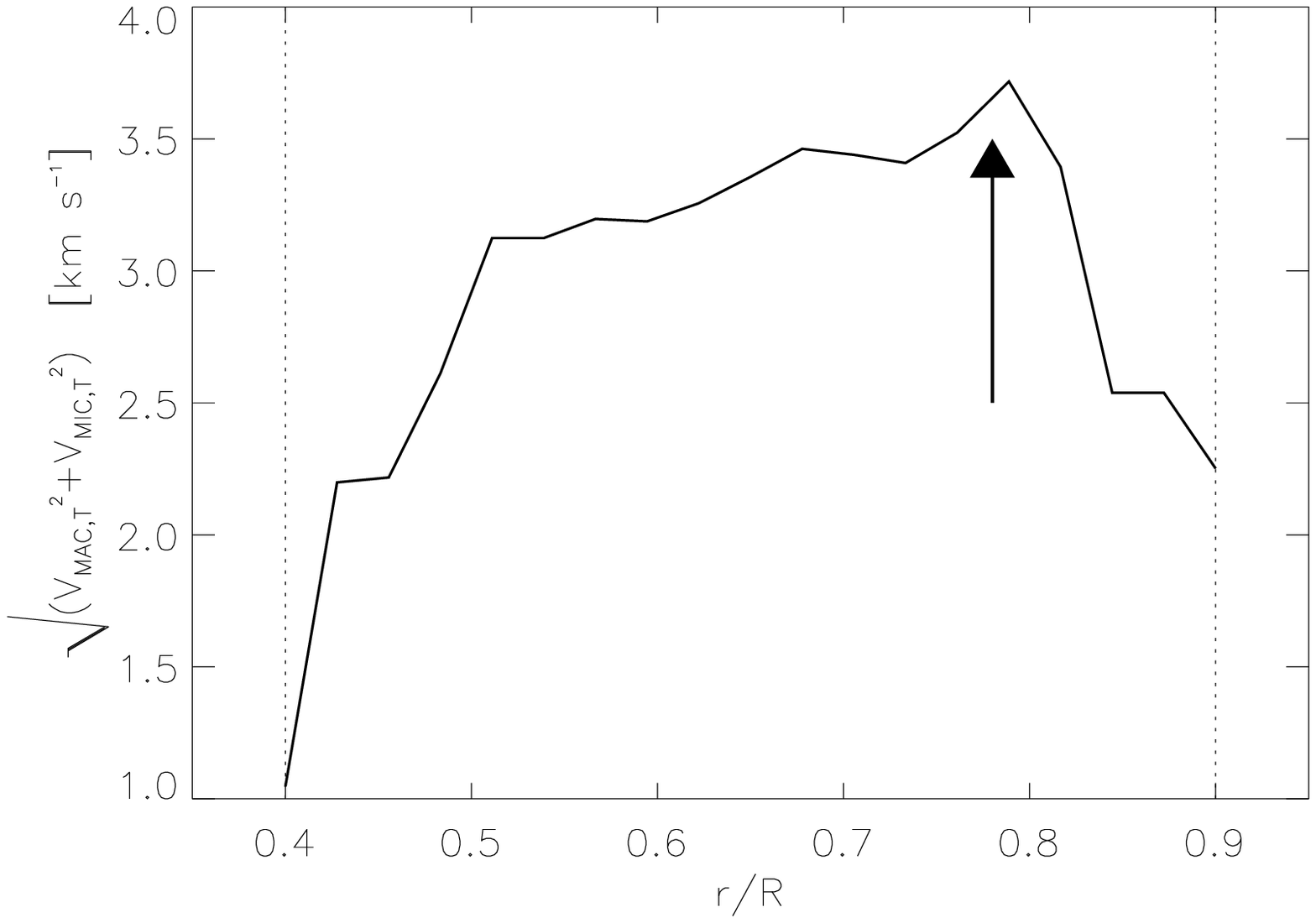}
\end{tabular}
\end{center}
\caption{Top left panel: Radial variation of the flow
 velocity inside the flux tube (solid line). Local
sound speed and tube's critical speed are also plotted
(dashed and dashed-dotted lines respectively). The vertical arrow
marks the position where the flow speed becomes supercritical, $r/R
\simeq 0.78$. Top right panel: radial variation of the equivalent width 
for Fe I 15648.5 \AA~ line. The open circles are three selected radial
positions before, during and after the shock occurs. Bottom left panels:
Intensity profiles corresponding to the 3 selected radial positions
(top right panel in this Figure) before (dashed line), during (solid
line) and after the shock (dotted line). Bottom right panels: radial 
variation of $\sqrt{v_{\rm mic,t}^2+v_{\rm mac,t}^2}$.}
\end{figure*}

\section{Summary and conclusions}%

We have presented the first full inversions of infrared penumbral 
spectropolarimetric data with a model that accounts for the vertical 
and horizontal inhomogeneities of the penumbral fine structure. This model is a
slightly modified version of the uncombed model of Solanki \& Montavon
(1993; cf Mart{\'\i}nez Pillet 2000) and allows for the presence of a
randomly orientated flux tube embedded in a surrounding
atmosphere that harbours a strong magnetic field. The main advantages of this
model, as compared to those used in
Paper I (see also Mathew et al. 2003; Bellot Rubio et al. 2003) are: 
{\bf a} -it contains two different 
atmospheres in the direction perpendicular to the observer (flux tube and
magnetic surrounding) which allows the observed
multilobed Stokes $V$ profiles to be easily reproduced; {\bf b} -these two atmospheres are also present
in the direction parallel to the observer line-of-sight and therefore the
model includes
gradients along the line of sight (in the form of sharp jumps in the physical 
quantities at the flux tube's boundaries) needed to produce 
asymmetric Stokes $V$ profiles ($\delta A \ne 0$); {\bf c} -flux tubes and their
magnetic surroundings are coupled to each other using total pressure balance at
all heights; {\bf d} -the current model requires fewer free parameters to
reproduce the data.

Feature {\bf b} makes this model suitable to investigate the vertical size of the
penumbral filaments, since the radius of the flux tube $R_t$, which is ultimately
linked to the amount of area asymmetry generated, is also obtained
from the inversion. The small amount of $\delta A$ shown by the Fe I lines at
1.56 $\mu$m is not enough to constrain the position of the flux tubes'
upper and lower boundaries, except at the outer penumbra where $\delta A$
is sufficiently large, allowing us to set a rather accurate position for
the lower boundary of the flux tube at around $\log\tau_5 \in [-0.5,0]$, while
the upper boundary remains undetected. Although for individual Stokes $V$ profiles the
area asymmetry produced by the uncombed model matches the observed
one only relatively inaccurately, general trends such as the radial behaviour
of $\delta A$ are reproduced fairly well.
We point out, in agreement with Paper I, that the use of lines that
show a much larger $\delta A$ should help in this matter (see
Borrero et al., in preparation).

Feature {\bf c} has allowed us to detect, for the first time, a strong radial decrease  
in the pressure difference between the flux tube and its surroundings that
is likely to induce an outward directed flow. This is to our mind the
strongest evidence so far supporting siphon flows (Meyer \& Schmidt 1968; Montesinos \&
Thomas 1993,1997) as the physical mechanism driving the Evershed effect.

In addition, we have seen that the Evershed flow already carries velocities as
large as $ v \sim 4$ km s$^{-1}$ in the inner penumbra. This has passed
unnoticed in previous works (see Schmidt \& Schlichenmaier 2000; Tritschler et
al. 2004) where only Stokes $I$ was considered. A possible explanation is
found in the small filling factor and inclinations, larger than 90 deg with respect to the
observer, of the magnetic field vector in the flow channels at the inner
penumbra. This geometry produces small line 
asymmetries in Stokes $I$, but large Stokes $V$ zero crossing shifts. At the
inner penumbra these fast flows are also associated with hot gas and flux
tubes that are somewhat
inclined with respect to the horizontal, in agreement with 
Schmidt \& Schlichenmaier (2000) and Rimmele (2004). These results are in close agreement with
the dynamical simulations of penumbral flux tubes by Schlichenmaier et
al. (1998a,1998b) and Schlichenmaier (2002). The tubes reach the 
same temperature as their surroundings very rapidly in the radial direction, and at the same 
time the flow speed increases smoothly (up to $v \sim 5$ km s$^{-1}$) 
as the pressure drops (Montesinos \& Thomas 1993, 1997).

At large radial distances the flow speed suffers a sudden decrease that is
associated with positions where the tubes return to the solar
interior (Westendorp Plaza et al. 1997). In addition to this well established
result, we have also detected a possible transition between critical and subcritical
velocities (as predicted by Montesinos \& Thomas) that is co-spatial with a rise in temperature 
and equivalent width at the outer penumbra, and seems to indicate that 
part of the kinetic energy is being dissipated into thermal energy. 
Therefore, part of the sudden drop in the velocity at the outer penumbra 
could be ascribed to the development of shock fronts.

\begin{acknowledgements}
We thank the referees, Drs. J. Thomas and B. Montesinos for valuable comments and suggestions.
\end{acknowledgements}

\end{document}